\documentclass[aps,prb,twocolumn,amsmath,amssymb,superscriptaddress,preprintnumbers,floatfix,eqsecnum,nofootinbib]{revtex4-2}

\usepackage{amsmath}
\usepackage{amsthm}
\usepackage{mathtools}
\usepackage{graphicx}
\usepackage{amssymb}
\usepackage{bm} 
\usepackage{ulem}   
\usepackage{slashed}
\usepackage{color}
\usepackage{bbm}
\usepackage{amsfonts}
\usepackage[breaklinks=true,colorlinks,citecolor=blue,linkcolor=blue,urlcolor=blue]{hyperref}
\usepackage{cases}
\usepackage{float}
\usepackage{url}
\usepackage{framed}
\usepackage{bbold}
\usepackage{cancel,soul,ulem}
\usepackage{appendix}
\usepackage{hhline}

\usepackage{xcolor}

\definecolor{siteorange}{HTML}{FF8900}
\definecolor{siteteal}{HTML}{229C89}
\definecolor{sitepurple}{HTML}{7D237A}
\definecolor{ringred}{HTML}{F04B53}

\definecolor{bleu}{rgb}{0.0, 0.5, 0.69}

\newcommand{\Z}{\mathbb{Z}}
\newcommand{\C}{\mathbb{C}}

\newcommand{\ket}[1]{\lvert #1 \rangle}

\def\=={\raisebox{0.35pt}{$\mathrm{:}$}\!\!=}

\definecolor{amar}{rgb}{0.9, 0.17, 0.31}
\definecolor{brred}{rgb}{0.8, 0.25, 0.33}
\definecolor{frblue}{rgb}{0.0, 0.45, 0.73}

\usepackage{tikz}
\usepackage{tikz-cd}
\usetikzlibrary{arrows}
\usetikzlibrary{intersections}
\usetikzlibrary{shapes.geometric}
\usetikzlibrary{decorations.pathmorphing, patterns,shapes}
\usetikzlibrary{decorations.markings}
\definecolor{siteorange}{HTML}{FF8900}
\definecolor{siteteal}{HTML}{229C89}
\definecolor{sitepurple}{HTML}{7D237A}

\usepackage[dvipsnames]{xcolor}

\newcommand{\w}{\omega_{N}}

\theoremstyle{definition}

\theoremstyle{remark}

\begin{document}	
\preprint{MIT-CTP/6149}

\title{
Invertible parafermionic topological phases in one dimension}
\author{Andr\'{e} L.G. Mudry}
\affiliation{
ETH Z\"urich, 8092 Z\"urich, Switzerland
}
\author{\"Omer M. Aksoy}
\affiliation{
Center for Theoretical Physics — a Leinweber Institute, Massachusetts Institute of Technology, Cambridge, MA 02139, USA
}

\begin{abstract}
Invertible topological phases of matter are described by gapped ground states
of local Hamiltonians that are non-degenerate on any closed spatial
manifold.  Classification of invertible topological phases realized by
bosonic or fermionic degrees of freedom is well-understood, particularly
in low dimensions. In one-dimensional space, the classification can be
understood through the boundary projective
representations. We extend this paradigm to one-dimensional invertible
topological phases of parafermionic degrees of freedom with
symmetries. We define cohomological invariants that characterize
distinct boundary projective representations and compute their
stacking rules. Among other results, we show (i) how these
cohomological invariants can be used to compute protected boundary
degeneracy and (ii) how, in contrast to bosons or fermions, invertible
phases of parafermions can form a non-Abelian group under stacking.
For order-$N$ parafermions with $N$ a prime power, we define
representative Hamiltonians realizing invertible topological phases
with $\Z^{3}_{N}$ symmetry.  We show that under stacking they generate a
non-Abelian group of order $2N^3$.  When $N$ is an odd prime, this
group is $\mathrm{Heis}(\Z_N)\rtimes \Z^{\,}_{2}$,
where $\mathrm{Heis}(\Z_N)$ is the
Heisenberg group of order~$N^{3}$.
\end{abstract}

\date{\today}

\maketitle

\tableofcontents

\section{Introduction}
\label{sec:Intro}

Although quantum matter is microscopically composed of electrons and ions
treated as point particles, its low-energy behavior is governed by emergent
quasiparticles. Examples include phonons, magnons, and excitons on the
bosonic side, and Landau quasiparticles on the fermionic side.
In three-dimensional space such point-like quasiparticles
only come in two kinds, bosons or fermions
\cite{Leinaas1977},
for which the many-body wavefunction is symmetric 
or antisymmetric under the exchange of identical excitations, respectively. 

In lower spatial dimensions, point-like excitations need not obey exclusively
bosonic or fermionic statistics, but may instead exhibit fractional
statistics~\cite{Leinaas1977,Wilczek1982}.
For example, point-like excitations in two-dimensional topologically
ordered phases, known as anyons, may obey either Abelian or
non-Abelian exchange statistics. In the Abelian case, exchanging two
identical anyons multiplies the many-body wavefunction by a phase that
need not be restricted to $\pm 1$~\cite{Halperin1984,Arovas1984}.
It has been proposed in
Refs.~\cite{Clarke2013,Lindner2012,Cheng2012,Vaezi2013} that
one-dimensional interfaces between two-dimensional topologically ordered states coupled by proximity effect to superconductors
can host degrees of freedom that provide
one-dimensional avatars of anyonic degrees of freedom. These so-called
order-$N$ \textit{parafermionic} degrees of freedom obey exchange
relations characterized by an $N$th root of unity.
Since then, various platforms have been proposed to host
parafermionic degrees of freedom~\cite{Mong2014,Clarke2014,Ebisu2017,Liang2019,Groenendijk2019,Cao2024,Liu2025,cao2026,Yoshitome2026,Bollmann2026,bollmann2026b}.
More recently, the experimental realizations~\cite{Cai2023,Zeng2023,Park2023,Xu2023,Ji2024,Redekop2024,Lu2024,Chen2024,Lu2025FCI} of fractional Chern insulators (FCIs)~\cite{Tang2011,Neupert2011,Sun2011,Sheng2011,Regnault2011}, 
open a new pathway to realize excitations
with fractionalized statistics. Unlike the earlier proposals,
FCIs do not require an external magnetic field to break time-reversal symmetry. This removes the tension between proximity superconductivity
and the magnetic field.

In one-dimensional space, order-$N$ parafermions can be thought of as
generalizations of Majorana fermions, the latter
corresponding to the case of $N=2$. In one-dimensional space, 
just as any Hamiltonian built out of Majorana operators
is related to a Hamiltonian 
built out of spin-1/2 via the Jordan-Wigner 
(JW) transformation~\cite{Jordan28}, any
Hamiltonian built out of order-$N$ 
parafermions is related to one
built out of $\Z^{\,}_{N}$ clock and shift 
operators via the Fradkin-Kadanoff (FK) 
transformation~\cite{Fradkin1980,Fendley2012}.

Motivated by proposals
for realizing parafermionic degrees of freedom
in one-dimensional systems, substantial attention has been devoted to
the phases of parafermionic Hamiltonians~\cite{Motruk2013,Bondesan2013,Fendley2014,Mong_2014b,Jermyn2014,
Li2015,Alicea2016,Alexandradinata2016,Iemini2017,Snizhko2018,Chew2018,Calzona2018,Schmidt2020}.
Of particular interest are invertible topological phases of
matter~\cite{Chen2013,Kapustin2015,Freed2021,Gaiotto2019}, whose local
Hamiltonians have a non-degenerate gapped ground state on any closed spatial
manifold.

An invertible topological phase is called \textit{trivial} if it admits
a representative whose non-degenerate ground state is a product of
localized states.
By contrast, under open boundary conditions, a Hamiltonian in a
\textit{non-trivial} invertible topological phase supports protected
degrees of freedom localized at its boundaries. These boundary degrees
of freedom give rise to a ground-state degeneracy in the thermodynamic
limit, with any finite-size splitting vanishing as the boundaries are
taken infinitely far apart.

We define a continuous deformation between two local Hamiltonians as a
continuous change of their short-range couplings, allowing also for the
addition or removal of decoupled local degrees of freedom in product
states.
Two Hamiltonians whose ground states realize invertible topological
phases are then said to be equivalent if they can be connected by such
a deformation without closing the spectral gap.

In addition to this notion of equivalence, invertible topological phases
are equipped with a composition law called \textit{stacking}, under
which they form a group. Stacking two phases amounts to combining their
degrees of freedom without coupling them. A representative Hamiltonian
for the resulting phase is given by
\begin{equation}
H_{\mathrm{stack}}^{\,}=
H_{1}^{\,}\otimes\mathbbm{1}^{\,}_{2}
+
\mathbbm{1}^{\,}_{1}\otimes H_{2}^{\,},
\end{equation}
where $H_{1}^{\,}$
and $H_{2}^{\,}$
are representative Hamiltonians for the two
phases. The trivial phase is the identity element under stacking, and
two phases are inverses of one another if their stacking yields the
trivial phase.

Additional symmetries can refine the classification of invertible
topological phases.\footnote{
For non-invertible symmetries, a symmetric product state need not
exist~\cite{Seifnashri2024,Pace2025}, 
in which case there is no conventional
notion of a trivial symmetric phase. Moreover, such symmetries do not
generally admit a stacking operation, and the corresponding
invertible topological phases therefore need not form a group. 
See Refs.~\cite{Pace2025,MengChengqi2024,Aksoy2025,Lu2025,You2026a,You2026b} for examples of invertible topological phases with non-invertible symmetries. In this
work, we consider only invertible symmetries for which both a trivial
symmetric phase and a stacking operation are well defined.}~In 
the presence of a symmetry, two phases are equivalent only if they
can be connected by a continuous deformation that neither closes the
spectral gap nor breaks the symmetry, explicitly or spontaneously.
An invertible topological phase that becomes trivial once the protecting
symmetry is relaxed is called a \textit{symmetry-protected topological}
(SPT) phase~\cite{Gu2009,Chen2013}.

The classification of invertible topological phases of bosons and fermions 
is well understood in one-dimensional space%
~\cite{Fidkowski2011,Chen2011a,Schuch2011,Bultinck2017,%
Turzillo2019,Bourne2021,Aksoy2022}.
In this work, we extend this paradigm to the invertible
phases of matter realized by Hamiltonians built out of order-$N$ 
parafermions in one-dimensional space, 
which we call
\textit{invertible parafermionic topological} (IPT) phases.
Following the strategy introduced in Ref.\ \cite{Aksoy2022},
we systematically classify IPT phases in one dimension
and derive explicitly 
their stacking rules given a global symmetry group $\mathcal{G}$. 
We briefly summarize our strategy and results in Sec.\ \ref{subsec:Summary}.

\subsection{Summary of the results}
\label{subsec:Summary}

\begin{table*}[t]
\caption{\label{tab:IPT table}
The groups of IPT phases of order-$N$ parafermions
$\mathrm{Inv}^{\mathrm{pf}}_N(\mathcal{G})$ 
and bosonic SPT phases 
$\mathrm{SPT}^{\mathrm{b}}(\mathcal{G})$ 
with symmetry $\mathcal{G}$.
We always consider $N$ to be a prime power, i.e., $N\equiv p^k$,
except for the last row, where $N$ is a composite
with the prime decomposition $N= N_1\times N_2 \times \cdots \times 
N_n$.
In each row, we compare the groups of IPT 
phases of order-$N$ parafermions
and SPT phases of boson, both with $\mathcal{G}$ symmetry.
We denote these groups by $\mathrm{Inv}^{\mathrm{pf}}_N(\mathcal{G})$
and $\mathrm{SPT}^{\mathrm{b}}(\mathcal{G})$, respectively.
When the group $\mathcal{G}$ is used for
bosonic SPT phases, we treat the subgroup $\Z^{\mathrm{pf}}_{N}$ as an
ordinary $\Z^{\,}_{N}$ symmetry.
When $N$ is a prime power, 
the IPT phases always have the structure 
$\mathrm{Inv}^{\mathrm{pf}}_N(\mathcal{G})
= \mathrm{SPT}^{\mathrm{pf}}_N(\mathcal{G})\rtimes\Z^{\,}_{2}$,
where $\mathrm{SPT}^{\mathrm{pf}}_N(\mathcal{G})$ 
denotes the SPT phases of order-$N$
parafermions.
Using the FK duality, the SPT phases of parafermions can be
mapped to the corresponding SPT phases of bosons. 
While the two groups have the same cardinality $|\mathrm{SPT}^{\mathrm{pf}}_N(\mathcal{G})| = |\mathrm{SPT}^{\mathrm{b}_N(\mathcal{G})}|$
they are different groups as the stacking rules differ.
In the second to last row, we take $p$ to be an odd integer 
$p=2l-1$ with $l=1,2,\cdots$. 
Therein, we denote by  $\mathrm{Heis}(\Z^{\,}_{N})$
the non-Abelian Heisenberg group of order $N^3$.}
\begin{center}
\begin{tabular}{cccc}
\hline\hline
$N$ \qquad&\qquad  $\mathcal{G}$ \qquad&\qquad 
$\mathrm{Inv}^{\mathrm{pf}}_N(\mathcal{G})$ 
\qquad&\qquad
$\mathrm{SPT}^{\mathrm{b}}(\mathcal{G})$
\\\hline
Any $(p,k)$ \qquad&\qquad $\Z^{\mathrm{pf}}_{N}$ \qquad&\qquad  $\Z^{\,}_{2}$ \qquad&\qquad $\Z_1$ 
\\\hline
Any $(p,k)$ \qquad&\qquad  $\Z^{\,}_{N}\times \Z^{\mathrm{pf}}_{N}$ \qquad&\qquad  
$\Z^{\,}_{N}\rtimes \Z^{\,}_{2}$ \qquad&\qquad $\Z^{\,}_{N}$ 
\\\hline
$(p,k)=(2,3)$ \qquad&\qquad  $\Z^{\mathrm{pf}}_{8}\rtimes (\Z^{\,}_{2} \times \Z^{\,}_{2})$ 
\qquad&\qquad  
$\Z^{3}_{2}$ \qquad&\qquad $\Z^2_2$ 
\\\hline
$(p,k)=(2l-1,k)$ \qquad&\qquad $\Z^{\,}_{N}\times \Z^{\,}_{N} \times \Z^{\mathrm{pf}}_{N}$ \qquad&\qquad  
$\mathrm{Heis}(\Z^{\,}_{N})\rtimes \Z^{\,}_{2}$ \qquad&\qquad $\Z^3_N$ 
\\\hline
$N= N_1\times N_2 \times \cdots \times N_n$ \qquad&\qquad  $\Z^{\mathrm{pf}}_{N}$ 
\qquad&\qquad  
$\Z^{n}_{2}$ \qquad&\qquad $\Z^{\,}_1$ 
\\
\hline&\\
\end{tabular}
\end{center}
\end{table*}

Throughout this paper, space is one-dimensional.
We consider local Hamiltonians built out of
order-$N$ parafermionic operators
that realize the parafermion algebra $\mathfrak{P}^{2L}_{N}$
[see Eq.\ \eqref{eq:alg 2L PFs}]. We
consider parafermions of prime-power order,
i.e., $N=p^{k}$ with $p$ a prime and $k$ a positive integer.
We discuss in Sec.\ \ref{subsec:general N}, a generalization to
the case of arbitrary integer $N$.

We argue that any local Hamiltonian realizing IPT phases 
of order-$N$ parafermions must have a global $\Z^{\mathrm{pf}}_{N}$ symmetry 
(Sec.\ \ref{subsec:PF Charge}), 
whose eigenvalues label the \textit{parafermion charge} of the states
in the global Hilbert space. This is a generalization of the $\Z^{f}_{2}$
fermion parity symmetry. Any additional symmetry group,
which we denote by $G^{\,}_{b}$,
is extended by $\Z^{\mathrm{pf}}_{N}$ to give rise to the global 
parafermionic symmetry group $\mathcal{G}$, such that 
$\Z^{\mathrm{pf}}_{N}\subset\mathcal{G}$ is a normal subgroup.

Following Ref.\ \cite{Aksoy2022},
we posit that,
for any Hamiltonian realizing an IPT phase
under open boundary conditions,
there exists a set of protected
boundary degrees of freedom such that
(i) they  form an algebra of order-$N$ parafermions
$\mathfrak{P}^{n}_{N}$ for some integer $n$,
(ii) and they render the ground state degenerate
in the thermodynamic limit.
If so, the distinct IPT phases are characterized by inequivalent 
boundary projective representations of the global symmetry group 
$\mathcal{G}$ on the algebra $\mathfrak{P}^{n}_{N}$. To classify 
such boundary projective representations we introduce a 
triplet of cohomological indices 
(Sec.\ \ref{sec:Boundary projective representations})
\begin{align}
\label{eq:triplet indices summary}
\big(\mu, (\kappa,\nu)\big)\in
\Z^{\,}_{2}\times
C^{1}\big(G^{\,}_{b},\Z^{\,}_{N}\big)\times
C^{2}\big(G^{\,}_{b},U(1)\big).
\end{align}
The index $\mu=0,1$ denotes whether there is an even or odd number
of parafermionic degrees of freedom associated with the boundary 
(${\mu=n\mod 2}$). The case of $\mu=1$ describes
the IPT phase that is beyond SPTs,
i.e., the non-trivial IPT phase whose ground state cannot be 
deformed into the product state under any local deformation.
When the group $\mathcal{G}$ is a non-split extension of $G^{\,}_{b}$ by 
$\Z^{\mathrm{pf}}_{N}$ (see Appendix~\ref{appsec:construction of G}), 
then the only compatible IPT phases
are SPTs and the index $\mu$ always takes the value $\mu=0$.

The remaining pair of indices $(\kappa,\nu)$ characterizes
the SPT portion of the IPT phases.
More specifically, the index $\kappa$ is related to 
the parafermion charge of the boundary representations of elements in 
$G^{\,}_{b}$, while the index $\nu$ encodes the projective representations 
of the elements in $G^{\,}_{b}$. The two indices $\nu$ and 
$\kappa$ correspond to the \textit{bosonic} part and the 
\textit{intrinsically} parafermionic part of the SPTs realized by order-$N$ parafermions. 

We derive the stacking rules (Sec.\ \ref{sec: Parafermionic Stacking Rules})
of IPT phases with $\mathcal{G}$ symmetry
by considering two boundary 
projective representations associated with indices 
$\big(\mu^{\,}_{1},(\kappa^{\,}_{1},\nu^{\,}_{1})\big)$
and
$\big(\mu^{\,}_{2},(\kappa^{\,}_{2},\nu^{\,}_{2})\big)$
and constructing \textit{stacked} indices
\begin{align}
\big(\mu^{\,}_{1\wedge 2}, (\kappa^{\,}_{1\wedge 2},\nu^{\,}_{1\wedge 2})\big)=
\big(\mu^{\,}_{1}, (\kappa^{\,}_{1},\nu^{\,}_{1})\big)
\wedge
\big(\mu^{\,}_{2}, (\kappa^{\,}_{2},\nu^{\,}_{2})\big),
\end{align}
by explicitly constructing the stacked
representation that acts on the degrees of freedom of
the parafermion algebras
$\mathfrak{P}^{n^{\,}_{1}}_{N}$
and
$\mathfrak{P}^{n^{\,}_{2}}_{N}$
in the same way as
the projective representations associated with the indices
$\big(\mu^{\,}_{1},(\kappa^{\,}_{1},\nu^{\,}_{1})\big)$
and
$\big(\mu^{\,}_{2},(\kappa^{\,}_{2},\nu^{\,}_{2})\big)$,
respectively.
We find that the stacked indices are given by
\begin{subequations}
\begin{align}
\label{eq:stacking rules combined summary}
\begin{split}
&
\mu^{\,}_{1\wedge2}=
\mu^{\,}_{1}
+
\mu^{\,}_{2}\ \mod\ 2,
\\
&
\kappa^{\,}_{1\wedge2}=
\kappa^{\,}_{1}
+
(-1)^{\mu^{\,}_{1}}\,
\kappa^{\,}_{2},
\\
&
\nu^{\,}_{1\wedge2}=
\nu^{\,}_{1} +\nu^{\,}_{2}
+
(-1)^{\mu^{\,}_{1}+1}\,
\frac{2\pi}{N}\kappa^{\,}_{1}\cup\kappa^{\,}_{2}.
\end{split}
\end{align}
Here, $\cup$ denotes the cup product of the
$\Z^{\,}_{N}$-valued 1-cochains $\kappa^{\,}_{1}$ and $\kappa^{\,}_{2}$,
which produces a $\Z^{\,}_{N}$-valued 2-cochain, as is explained in
Appendix~\ref{appsec:Group Cohomology}.
Remarkably, these stacking rules are not Abelian in that
\begin{align}
\big(\mu^{\,}_{1\wedge 2}, (\kappa^{\,}_{1\wedge 2},\nu^{\,}_{1\wedge 2})\big)\neq
\big(\mu^{\,}_{2\wedge 1}, (\kappa^{\,}_{2\wedge 1},\nu^{\,}_{2\wedge 1})\big).
\end{align}
\end{subequations} 
Physically, this non-commutativity stems from the fact that
parafermionic degrees of freedom require a total ordering and are
therefore intrinsically one-dimensional. When stacking two IPT phases,
one must consequently prescribe the relative ordering of the
parafermionic degrees of freedom belonging to the two phases.

In the stacking rules \eqref{eq:stacking rules combined summary},
there are two sources of non-commutativity. The first 
arises from the factor
$(-1)^{\mu^{\,}_{1}}$. The second
arises from the fact that
\begin{align}
\label{eq:second source of noncomm}
\kappa^{\,}_{1}\cup\kappa^{\,}_{2} \neq
\kappa^{\,}_{2}\cup\kappa^{\,}_{1},
\end{align}
in general.
We denote the group of IPT phases of order-$N$ 
parafermions with $\mathcal{G}$ by 
$\mathrm{Inv}^{\mathrm{pf}}_N(\mathcal{G})$
and the group of SPT phases by $\mathrm{SPT}^{\mathrm{pf}}_N(\mathcal{G})$.
We find the relation
\begin{align}
\mathrm{Inv}^{\mathrm{pf}}_N(\mathcal{G})=
\mathrm{SPT}^{\mathrm{pf}}_N(\mathcal{G})\rtimes\Z^{\,}_{2}.
\end{align}
The semidirect product by $\Z^{\,}_{2}$ is due to the factor of
$(-1)^{\mu^{\,}_{1}}$.
Eq.\ \eqref{eq:second source of noncomm} leads to the group
$\mathrm{SPT}^{\mathrm{pf}}_N(\mathcal{G})$ of SPT phases 
being non-Abelian in general. 
The fact that 
a semidirect product by $\Z^{\,}_{2}$ arises when stacking
IPT phases was first pointed out in Ref.\ \cite{Motruk2013}.
We extend this by showing that even the group of SPT phases of parafermions
can be non-Abelian.
In Table \ref{tab:IPT table}, we compute the classification
of IPT phases of order-$N$ parafermions with a given symmetry group 
$\mathcal{G}$. We contrast this with the classification of bosonic 
SPT phases with $\mathcal{G}$ symmetry~\footnote{For bosonic 
SPTs, the $\Z^{\mathrm{pf}}_N\subset \mathcal{G}$ subgroup is 
treated as an ordinary $\Z_N$ symmetry.}. We note that
the group of SPT phases of parafermions $\mathrm{SPT}^{\mathrm{pf}}_N(\mathcal{G})$ 
is always of the same order as those of bosons. However, 
they differ in their group structure which can be non-Abelian.

In the last row of Table \ref{tab:IPT table}, we consider 
IPT phases for parafermions of composite order 
$N= N_1\times N_2 \times \cdots \times N_n$, with no additional
symmetry. Interestingly, in this case there is a distinct 
IPT phase that is not an SPT for each prime factor (also see Sec.\
\ref{subsec:general N}).

Finally, we explicitly construct (Sec.\ \ref{sec:stacking example})
lattice Hamiltonians that realize IPT phases of order-$N$ parafermions
with $\Z^{\,}_{N}\times \Z^{\,}_{N} \times \Z^{\mathrm{pf}}_{N}$ 
symmetry (with $N$ odd prime power). 
We show that there are $2\,N^3$ IPT phases for each of which
the Hamiltonians can be generated by stacking the four Hamiltonians
defined in Eq.\ \eqref{eqs: generating hamiltonians}. The 
group of IPT phases is then given by $\mathrm{Heis}(\Z_N)\rtimes \Z_2$
where $\mathrm{Heis}(\Z_N)$ is the non-Abelian Heisenberg group of
order $N^3$. The group of SPT phases is given by the subgroup
$\mathrm{Heis}(\Z_N)$ which is non-Abelian as opposed to the 
case of bosons or fermions.

\subsection{Outline}
\label{subsec:Outline}

The rest of the paper is organized as follows.  
In Sec.\ \ref{sec:Defs}, we define the order-$N$ parafermion algebra,
the Hilbert space on which the parafermionic operators
act and the protecting symmetries.
In Sec.\ \ref{sec:Boundary projective representations}, we consider 
the projective representations realized at the boundaries of IPT phases
and introduce cohomological indices that classify them. 
Section \ref{sec: Parafermionic Stacking Rules} presents a derivation 
of the stacking rules of IPT phases, which,
unlike invertible topological phases of bosons and fermions,  
can form
non-Abelian groups under stacking.
In Sec.\ 
\ref{sec:stacking example}
we provide an explicit example of Hamiltonians
that realize IPT phases with a $\mathcal{G}\cong\Z^{3}_{N}$ symmetry.
These phases form a non-Abelian group of order
$2N^{3}$ under stacking
(the corresponding SPT phases form a non-Abelian group of order $N^{3}$
under stacking).
Sec.\ \ref{sec:conclusion} concludes the paper.
Appendices 
\ref{appsec:Group Cohomology},
\ref{appsec:construction of G}, 
and
\ref{appsec:odd parafermion rep identities}
provide summaries of
group cohomology,
properties of symmetry group $\mathcal{G}$,
and proofs of Eq.\ \eqref{eq:general proj rep mu odd}, respectively.

\section{Definitions and Assumptions}
\label{sec:Defs}

We consider a one-dimensional lattice of $L$ sites, each of which
supports a pair of order-$N$ parafermionic degrees of freedom. 
For simplicity, we focus on parafermions of prime-power order, i.e., $N=p^{k}$
with $p$ a prime and $k$ a positive integer.
The case of general $N$ is discussed in Sec.~\ref{subsec:general N}.

Without loss of generality, we associate an $N$-dimensional local
Hilbert space $\mathcal{H}^{\,}_{j}$
to each site $j=1,\cdots,L$ of the lattice.
Thus, for a lattice of length $L$,
the total Hilbert space
$\mathcal{H}$ is of dimensionality $N^{L}$.
We define $2L$ unitary parafermion operators
\begin{subequations}
\label{eq:alg 2L PFs}
\begin{align}
\Big\{
\gamma^{\,}_{1},\,
\gamma^{\,}_{2},\,
\cdots,\,
\gamma^{\,}_{2L-1},\,
\gamma^{\,}_{2L}
\Big\}
\end{align}
acting on the global Hilbert space
$\mathcal{H}$
such that they satisfy the algebra
\begin{align}
\label{eq:Order N Pf Algebra}
\gamma^{N}_{j}=1,
\quad
\gamma^{\,}_{j}\,
\gamma^{\,}_{k}=
\omega^{\mathrm{sgn}(k-j)}_{N}\,
\gamma^{\,}_{k}\,
\gamma^{\,}_{j},
\quad
\omega^{\,}_{N}:=
e^{\mathrm{i}\frac{2\pi}{N}},
\end{align}
\end{subequations}
where $\mathrm{sgn}(x)$ is the signum function that is $0$ when ${x=0}$ 
and $x/|x|$ otherwise. We denote by $\mathfrak{P}^{2L}_{N}$ 
the algebra generated by the $2L$ order-$N$ parafermions obeying Eq.\ 
\eqref{eq:Order N Pf Algebra}.

To define a local basis of the $N^{L}$-dimensional global Hilbert space
$\mathcal{H}$, it
is convenient to pair up the $2L$ parafermion operators
using the notation
\begin{subequations}
\label{eq:Pf local charge}
\begin{equation}
\xi^{\,}_{j}\equiv
\gamma^{\,}_{2j-1},
\qquad
\eta^{\,}_{j}\equiv
\gamma^{\,}_{2j},
\label{eq:Pf local charge a}
\end{equation}
for ${j=1,\,2,\,\cdots,\,L}$. We can then define
for each site $j$ the pairwise commuting local operators 
\begin{equation}
Q^{\,}_{j}=
e^{\mathrm{i}\pi\frac{N+1}{N}}\,
\xi^{\,}_{j}\,
\eta^{\dag}_{j},
\qquad
[Q^{\,}_{j},\, Q^{\,}_{k}] = 0,
\qquad
Q^{N}_{j}=1.
\label{eq:Pf local charge b}
\end{equation}
An orthonormal basis spanning the global Hilbert space
$\mathcal{H}$
is constructed from the simultaneous eigenstates 
\begin{equation}
{\ket{\bm{\alpha}}\equiv
\ket{\alpha^{\,}_{1},\,\alpha^{\,}_{2},\,\cdots,\, \alpha^{\,}_{L}}}
\label{eq:Pf local charge c}
\end{equation}
of the $Q^{\,}_{j}$ operators, such that
\begin{align}
Q^{\,}_{j}\,
\ket{\bm{\alpha}}=
\omega^{\alpha^{\,}_{j}}_N\,
\ket{\bm{\alpha}},
\label{eq:Pf local charge d}
\end{align}
\end{subequations}
where ${\alpha^{\,}_{j}= 0,\,1,\,\cdots,\, N-1}$
and $j=1,\cdots,L$.

\subsection{Parafermion charge symmetry}
\label{subsec:PF Charge}

From the local operators \eqref{eq:Pf local charge b}
we construct the global operator
\begin{subequations}
\label{eq:global Pf charge op}
\begin{align}
Q:=
\prod_{j=1}^{L}Q^{\,}_{j},
\qquad
Q\,\ket{\bm{\alpha}}=
(\omega^{\,}_N)^{\sum\limits_{j=1}^{L} \alpha^{\,}_{j}}\,
\ket{\bm{\alpha}}.
\end{align}
The operator $Q$
is thus diagonal in the basis
(\ref{eq:Pf local charge c})
and its eigenstate $\ket{\bm{\alpha}}$ has the eigenvalue
$\omega_{N}^{q^{\,}_{\bm{\alpha}}}$, where
\begin{align}
q^{\,}_{\bm{\alpha}}=\sum_{j=1}^{L}\alpha^{\,}_{j} \mod N.
\end{align}
Moreover, conjugation of any individual parafermion 
$\gamma^{\,}_{j}$ by $Q$ gives
\begin{align}
Q\,\gamma^{\,}_{j}\,Q^{\dag}=
\omega^{\,}_{N}\,
\gamma^{\,}_{j},
\qquad
j=1,\cdots,2L.
\end{align}
\end{subequations}
Hence, we call $Q$ the global \textit{parafermion charge operator}.
A basis state $\ket{\bm{\alpha}}$ carries the parafermion charge $q^{\,}_{\bm{\alpha}}\! \mod N$, while individual parafermion operators $\gamma_j$ carry unit parafermion charge.
The parafermion charge operator \eqref{eq:global Pf charge op}
generalizes global fermion parity, which is recovered when ${N=2}$.
One can show that such an operator can only exist if there is an even total number of parafermion operators.

Any IPT phase built out of order-$N$ parafermions must have a 
$\Z^{\mathrm{pf}}_N$ parafermion charge symmetry implemented by $Q$.
On the one hand, just like fermion parity, $\Z^{\mathrm{pf}}_{N}$ symmetry
is more than an ordinary $\Z^{\,}_{N}$ symmetry,
as it dictates the mutual statistics obeyed by local operators.
On the other hand, unlike fermion parity,
$\Z^{\mathrm{pf}}_{N}$ symmetry can be broken,
either spontaneously or explicitly.

To appreciate why $\Z^{\mathrm{pf}}_{N}$ symmetry
is more than an ordinary $\Z^{\,}_{N}$ symmetry,
consider two operators $A^{\,}_{j}$ and $B^{\,}_{k}$
from $\mathfrak{P}^{2L}_{N}$ with definite parafermion charges 
$q^{\,}_{A}$ and $q^{\,}_{B}$, respectively.
Assume that $A^{\,}_{j}$ and $B^{\,}_{k}$
only act non-trivially
on the finite non-overlapping intervals $I^{\,}_{j}$
and $I^{\,}_{k}$ centered around sites $j$ and $k$,
respectively. 
It then follows that
\begin{equation}
A^{\,}_{j}\,
B^{\,}_{k}=
\omega^{q^{\,}_{A}\,q^{\,}_{B}\,\mathrm{sgn}(k-j)}_{N}\,
B^{\,}_{k}\,
A^{\,}_{j}.
\label{eq:algebra well separated Pf ops}
\end{equation}
Importantly, the algebra \eqref{eq:algebra well separated Pf ops}
holds irrespective of the separation between the sites $j$ and $k$.
This means that the two well-separated 
operators $A^{\,}_{j}$ and $B^{\,}_{k}$ can be 
simultaneously measured only when $q^{\,}_{A}\,q^{\,}_{B} = 0 \! \mod N$.

This observation can be applied to any
local Hamiltonian built out of parafermion operators
\begin{subequations}
\label{eq:local Hamiltonian}
\begin{align}
H = \sum_{j}
h_j,
\label{eq:local Hamiltonian a}
\end{align}
where each summand $h_j$ acts non-trivially only on an interval centered around site $j$ containing no more than $r$ sites.
The local energy density is then a well-defined local observable if and only 
if  the two Hamiltonian terms $h_j$ and $h_k$ can be simultaneously measured  
for any $j$ and $k$ that are sufficiently separated, i.e., 
\begin{align}
\label{eq:local Hamiltonian b}
[h_j,h_k] = 0
\end{align}
\end{subequations}
for any $j,k$ such that $|j-k|>2r$.
The algebra \eqref{eq:algebra well separated Pf ops} then implies that 
if each $h_j$ has vanishing parafermion charge, i.e., $[Q,H]=0$, then
the local energy density becomes a well-defined observable.
In other words, the condition \eqref{eq:local Hamiltonian b} is satisfied
when Hamiltonian \eqref{eq:local Hamiltonian a} 
is $\Z^{\mathrm{pf}}_N$ symmetric. 
The converse is not true, i.e.,
the condition \eqref{eq:local Hamiltonian b} can be satisfied without the 
full $\Z^{\mathrm{pf}}_N$ symmetry.

Unlike fermion parity, the $\Z^{\mathrm{pf}}_{N}$
parafermion charge symmetry can be broken spontaneously,
e.g.,\ Refs.~\cite{Bondesan2013,Alexandradinata2016}. 
To see this, consider the
parafermion operators $\gamma^{p^{a}}_{j}$ with
the integer ${a \geq k/2}$ (recall that ${N\equiv p^k}$).
These operators obey the mutual bosonic statistics
\begin{align}
\gamma^{p^{a}}_{j}\,
\gamma^{p^{a}}_{k}=
\gamma^{p^{a}}_{k}\,
\gamma^{p^{a}}_{j}
\end{align}
for any ${j,k=1,\cdots,2L}$. In other words, 
such parafermionic operators can play the role of
local order parameters.
If such an operator gains a non-vanishing expectation value, 
it leads to the \textit{spontaneous} breaking of the parafermion 
charge symmetry from $\Z^{\mathrm{pf}}_{p^{k}}$ down to $\Z^{\mathrm{pf}}_{p^{a}}$.
The ground-state manifold is then $p^{k-a}$-fold degenerate and 
does not realize an invertible phase of matter.

Unlike fermion parity, the $\Z^{\mathrm{pf}}_{N}$
parafermion charge symmetry can also be \textit{explicitly} broken.
To see this, consider the terms $\gamma^{p^{a}}_{j}$ with 
$a\geq k/2$, which are local operators that
commute on distinct sites, i.e., they are local bosons.
Adding such a term to a Hamiltonian breaks explicitly the 
$\Z^{\mathrm{pf}}_{p^{k}}$ down to $\Z^{\mathrm{pf}}_{p^{a}}$
symmetry. More generally, one can add any local operator of charge
$p^a$ which results in a explicit breaking of the parafermion charge
symmetry. We then expect the problem of classifying 
IPT phases realized in the presence of such terms
to reduce to that with $\Z^{\mathrm{pf}}_{p^{a}}$ symmetry. 

Thus, in our classification of IPT phases of 
order-$N$ parafermions, we will always assume that the
$\Z^{\mathrm{pf}}_{N}$ parafermion charge symmetry is present and not broken,
neither spontaneously nor explicitly.

\subsection{Parafermionic symmetry groups}
\label{sec:def Gpf}

An invertible parafermionic phase of matter can have more
symmetries than just the $\Z^{\mathrm{pf}}_{N}$ parafermion charge symmetry.
We denote by $G^{\,}_{b}$ the group of unitary symmetries other than
$\Z^{\mathrm{pf}}_{N}$. The elements of the group $G^{\,}_{b}$ are also called
bosonic symmetries in the literature~\cite{Turzillo2019}, hence the subscript $\textit{b}$.

Demanding that the additional symmetries in group $G_b$ 
are locality preserving implies that every symmetry transformation 
must preserve the $\Z^{\mathrm{pf}}_N$ grading, i.e.,
under a symmetry transformation a definite parafermion charge sector
must be mapped to another definite parafermion charge sector.
This means that the $\Z^{\mathrm{pf}}_N$ symmetry must be a normal
subgroup of the total symmetry group $\mathcal{G}$, see Appendix \ref{appsubsec:Zn normal subgroup}.
The group $\mathcal{G}$ is 
then given as a group extension of
$G^{\,}_{b}$ by $\Z^{\mathrm{pf}}_{N}$ specified by the short exact
sequence (see Appendix \ref{appsubsec:G as an extension})
\begin{align}
1\to
\Z^{\mathrm{pf}}_{N}\underset{\iota}{\to}
\mathcal{G}\underset{\pi}{ \to}
G^{\,}_{b} \to 1,
\end{align}
where $\iota$ is an injective inclusion map and $\pi$ a surjective
quotient map such that ${\mathrm{im}\,\iota=\mathrm{ker}\,\pi}$.
The group isomorphism theorems then imply that $\mathrm{im}\,\iota$ is a
normal subgroup and
$G^{\,}_{b}\cong
\mathcal{G}/\mathrm{im}\,\iota\equiv
\mathcal{G}/\Z^{\mathrm{pf}}_{N}$.
Given a group $G^{\,}_{b}$,
a general group extension is specified by two pieces of data:
\begin{enumerate}
\item
a group action
$\rho_{(\cdot)}:G^{\,}_{b}\to\mathrm{Aut}(\Z^{\mathrm{pf}}_{N})$
of $G^{\,}_{b}$ on the parafermion charge symmetry $\Z^{\mathrm{pf}}_{N}$,
\item
and an extension class $[c]$ in the second cohomology group
$H^{2}\left(G^{\,}_{b},[\Z^{\mathrm{pf}}_{N}]^{\,}_{\rho}\right)$.
\end{enumerate}
We denote the elements of the total symmetry group $(\mathcal{G}, \circ)$
by $\tilde{g}$. 
We can alternatively represent an element 
$\tilde{g}\in\mathcal{G}$
as the pair $(z,g)$, where $z\in(\Z^{\mathrm{pf}}_{N}, +)$
and $g\in (G^{\,}_{b}, \cdot)$ with the respective group identity elements $0\in\Z^{\mathrm{pf}}_{N}$ and $e\in G^{\,}_{b}$.
In this representation, the group composition  of two elements $\tilde{g}\equiv(z,g)$
and
$\tilde{h}\equiv(u,h)$ in $\mathcal{G}$ becomes
\begin{align}
\label{eq:group composition in Gpf}
(z,g)\circ(u,h)=\big(z+\rho_{g}\,u+c(g,h),\,g\,h\big),
\end{align}
where the 2-cochain $c$ is a representative of the extension class $[c]$.
Note that the group action $\rho$ on $\Z^{\mathrm{pf}}_{N}$
reduces to the multiplication by an integer modulo $N$, whence we write
$\rho_{g}(z)\equiv\rho_{g}\,z$.

The parafermion charge symmetry $\Z^{\mathrm{pf}}_{N}$
can be viewed as a normal subgroup of $\mathcal{G}$
such that
\begin{equation}
G^{\,}_{b}\cong\mathcal{G}/\Z^{\mathrm{pf}}_{N}.
\end{equation}  
When the group action $\rho$ is trivial, i.e.,
$\rho_{g}=1$ for all $g\in G^{\,}_{b}$,
$\Z^{\mathrm{pf}}_{N}$ lies in the center of $\mathcal{G}$.

The possible group actions $\rho$ are severely restricted.
As shown in Appendix \ref{appsubsec:group actions are involutive}
every automorphism in the image of $\rho$ must be
involutions. For $N=p^{k}$, the group of involutive automorphisms of
$\Z^{\mathrm{pf}}_{p^{k}}$ is
\begin{align}
\label{eq:involutive group}
\mathcal{I}\left(\Z^{\mathrm{pf}}_{p^{k}}\right)
=
\begin{cases}
\Z^{\,}_1, &\hbox{if $p=2$ and $k=1$,}\\
\Z^{\,}_{2}, &\hbox{if $p=2$ and $k=2$,}\\
\Z^{\,}_{2}\times\Z^{\,}_{2},
&\hbox{if $p=2$ and $k>2$,}\\
\Z^{\,}_{2}, &\hbox{if $p>2$.}
\end{cases}
\end{align}
The result for generic $N$ follows from
Eq.\ \eqref{eq:involutive group} by applying the Chinese
remainder theorem.

For any prime $p$ and positive integer $k$,
Eq.\ \eqref{eq:involutions equation} admits the solutions
$\rho^{\,}_{g}=\pm1$.
They generate a $\Z^{\,}_{2}$ subgroup, except when $p=2$ and $k=1$, where these two solutions
coincide modulo $2$.
When $p=2$ and $k>2$, there are two additional solutions,
$\rho^{\,}_{g}=2^{k-1}\pm1$.
Together, the four solutions form the group
$\Z^{\,}_{2}\times\Z^{\,}_{2}$.%
~\footnote{
Note that
$(2^{k-1}\pm1)^{2}=1\!\mod 2^{k}$
and
$(2^{k-1}+1)(2^{k-1}-1)=-1\!\mod 2^{k}$.
}  

Physically, the solution $\rho^{\,}_{g}=-1$ implements
\textit{charge conjugation},
as it maps an operator of parafermion
charge $q$ to one of charge $-q$.
Its action on an individual parafermion can be
represented as
\begin{align}
\gamma^{\,}_{j}
\mapsto
\gamma^{\dag}_{j}.
\end{align}

When $p=2$ and $k>2$, the two additional solutions
$\rho^{\,}_{g}=2^{k-1}\pm1$ define what we call
\textit{special charge conjugation symmetries}.
Consider a single parafermion operator $\gamma_j$
which has unit charge $q=1$.
The solution
$\rho^{\,}_{g}=2^{k-1}-1$
combines charge conjugation with the attachment of a parafermion
charge $2^{k-1}$.
By contrast, the solution
$\rho^{\,}_{g}=2^{k-1}+1$
attaches a parafermion charge $2^{k-1}$ without implementing
charge conjugation.

For Majorana fermions, corresponding to $N=2$, one has
$-1=1\!\mod 2$.
Consequently, charge conjugation acts trivially on the
$\Z^{\mathrm{pf}}_{2}$ charge.
A non-trivial action of charge conjugation on the parafermion
charge is therefore possible only for $N\geq3$.

\section{Boundary projective representations}
\label{sec:Boundary projective representations}

An invertible parafermionic topological (IPT) phase is realized by a
local Hamiltonian
\begin{subequations}
\label{eq:IPTP Ham}
\begin{align}
H=
\sum_{j=1}^{L}
h^{\,}_{j},
\label{eq:IPTP Ham a}
\end{align}
such that
\begin{enumerate}
\item
each local term $h^{\,}_{j}$
is built out of order-$N$ parafermion operators
and has finite support,

\item
the ground state is non-degenerate and gapped in the thermodynamic
limit $L\to\infty$,

\item
and $H$ is $\mathcal{G}$-invariant, i.e.,
\begin{equation}
[U(z,g),h^{\,}_{j}]=0,
\qquad
j=1,\cdots,L,
\label{eq:IPTP Ham b}
\end{equation}
where $U(z,g)\in\mathfrak{P}_{N}^{2L}$
is a unitary operator
and the set of such operators
labeled by $(z,g)\in\mathcal{G}$ realizes
a faithful representation
of $\mathcal{G}$.
\end{enumerate}
\end{subequations}
Since the ground state $\ket{\psi}$
of the Hamiltonian \eqref{eq:IPTP Ham a} is non-degenerate and symmetric,
the representation $U(z,g)$  
can be chosen to be linear, i.e.,
\begin{align}
\label{eq:bulk representation}
\begin{split}
&
U(z,g)\,U(u,h)=
U\big((z,g)\circ (u,h) \big),
\\
&
U(z,g)\,\ket{\psi}=
e^{\mathrm{i}\varphi^{\,}_{z,g}}\,\ket{\psi}, 
\quad \varphi^{\,}_{z,g}\in [0,2\pi).
\end{split}
\end{align}
When open boundary conditions are imposed, the terms $h^{\,}_{j}$
in Eq.\ \eqref{eq:IPTP Ham a}
that are simultaneously supported around the sites 
$j=1$ and $j=L$
will be absent. In that case, there will be operators
that can be associated with the left and 
right ends of the open chain that commute with the Hamiltonian. 
Such operators describe the  local degrees of freedom 
associated with the right and the left boundaries of the open chain.
Accordingly, we assume the existence of the sets
\begin{align}
\label{eq:bd parafermions}
\begin{split}
&
\mathcal{O}^{\,}_{\mathrm{L}}=
\left\{
\gamma^{\,}_{\mathrm{L},\,1}, \, 
\gamma^{\,}_{\mathrm{L},\,2},
\cdots,\,
\gamma^{\,}_{\mathrm{L},\,n^{\,}_{\mathrm{L}}}
\right\},
\\
&
\mathcal{O}^{\,}_{\mathrm{R}}=
\left\{
\gamma^{\,}_{\mathrm{R},\,1}, \, 
\gamma^{\,}_{\mathrm{R},\,2},
\cdots,\,
\gamma^{\,}_{\mathrm{R},\,n^{\,}_{\mathrm{R}}}
\right\},
\end{split}
\end{align}
that are associated with the left and right boundaries, respectively.
The operators in the sets $\mathcal{O}^{\,}_{\mathrm{L}}$
and $\mathcal{O}^{\,}_{\mathrm{R}}$ satisfy the order-$N$ parafermion 
algebra \eqref{eq:Order N Pf Algebra} and, thus,
generate the parafermion algebras
$\mathfrak{P}^{n^{\,}_{\mathrm{L}}}_{N}$ and $\mathfrak{P}^{n^{\,}_{\mathrm{R}}}_{N}$,
respectively.
By assumption, every operator in the set  $\mathcal{O}^{\,}_{\mathrm{R}}$ 
lies to the right of those in the set $\mathcal{O}^{\,}_{\mathrm{L}}$, 
whence the union $\mathcal{O}^{\,}_{\mathrm{L}}\sqcup\mathcal{O}^{\,}_{\mathrm{R}}$
generates the parafermion algebra
$\mathfrak{P}^{n^{\,}_{\mathrm{L}}+n^{\,}_{\mathrm{R}}}_{N}$.

We demand that the degrees of freedom on the left and right boundaries 
together act on a well-defined global Hilbert space.%
~\footnote{
See Sec.\ \ref{subsec:defs indices odd} for 
a more detailed discussion on why an even number of parafermions 
is needed to define the Hilbert space.
} 
In other words, 
the total number of boundary parafermion operators must be even, i.e.,
\begin{align}
n^{\,}_{\mathrm{L}}+n^{\,}_{\mathrm{R}}=0 \mod 2.
\label{eq:parafermion number consistency equation}
\end{align}

We also demand that the boundary algebras generated 
by the sets of operators \eqref{eq:bd parafermions} 
are compatible with the 
bulk symmetries $U(\tilde{g})$ in the sense that 
\begin{align}
U(\tilde{g})\,
\mathfrak{P}^{n^{\,}_{\mathrm{B}}}_{N}\,
U^{\dag}(\tilde{g})=
\mathfrak{P}^{n^{\,}_{\mathrm{B}}}_{N}
\label{eq:bd compatibility cond}
\end{align}
for all elements $\tilde{g}\in \mathcal{G}$ and ${\mathrm{B}=\mathrm{L,\,R}}$.
A boundary that does not satisfy this
compatibility condition explicitly breaks the $\mathcal{G}$ symmetry which
we do not consider.

The bulk symmetry operators \eqref{eq:bulk representation} induce an
action on the boundary degrees of freedom through conjugation. We assume that the boundary
degrees of freedom \eqref{eq:bd parafermions} from the set
$\mathcal{O}_{\mathrm{B}}$ realize a \textit{faithful} representation
$U^{\,}_{\mathrm{B}}$ of the symmetry group $\mathcal{G}$
such that%
~\footnote{
One may need to add ``ancilla'' degrees of freedom
to the set $\mathcal{O}_{\mathrm{B}}$ of boundary operators for it to
be sufficiently large to accommodate a faithful representation of
$\mathcal{G}$. The gapped boundary degrees of freedom is a natural
source of such ancillas. See Sec.\ \ref{sec:stacking example} for an
example.
}
\begin{align}
\label{eq:bd rep key relation}
U(\tilde{g})\,
\gamma^{\,}_{\mathrm{B},\,j}\,
U^{\dag} (\tilde{g})=
U^{\,}_{\mathrm{B}}(\tilde{g})\,
\gamma^{\,}_{\mathrm{B},\,j}\,
U^{\dag}_{\mathrm{B}}(\tilde{g}),
\end{align}
for all $\tilde{g}\in\mathcal{G}$ and $j=1,\cdots,n^{\,}_{\mathrm{B}}$.
Physically, the condition \eqref{eq:bd rep key relation} means that the 
boundary representation $U^{\,}_{\mathrm{B}}(\tilde{g})$
implements the same transformation
rules as the bulk representation $U(\tilde{g})$. In practice, 
Eq.\ \eqref{eq:bd rep key relation} is the key relation that
we use to explicitly construct the boundary representations of
$\mathcal{G}$.

While the bulk representation \eqref{eq:bulk representation} is 
linear by definition, the boundary symmetry operators
$U^{\,}_{\mathrm{B}}(z,g)$
may realize a non-trivial projective representation. 
Distinct IPT phases 
are characterized by distinct equivalence classes of boundary projective representations. 
In what follows, we introduce cohomological indices 
$(\mu, [(\kappa, \nu)])$ that 
characterize the boundary projective representations that 
can be realized by order-$N$ parafermions given the group $G^{\,}_{b}$.
Without loss of generality, we will consider
the degrees of freedom on a single boundary $\mathrm{B}$.
We will treat the cases of even and odd 
$n^{\,}_{\mathrm{B}}$ separately.  
To distinguish these, we introduce a $\Z^{\,}_{2}$-valued index 
$\mu$ whose values $\mu=0,1$ correspond to the case of an even and 
odd number of boundary parafermions, respectively. 

\subsection{Even number of boundary parafermions}
\label{subsec:Boundary projective rep even}

When the number of boundary parafermion operators is even 
($\mu=0$), the parafermionic operators 
from the set $\mathcal{O}^{\,}_{\mathrm{B}}$ act on a local
Hilbert space $\mathcal{H}^{\,}_{\mathrm{B}}$ associated with the
boundary $\mathrm{B}$. We can label the states in 
$\mathcal{H}^{\,}_{\mathrm{B}}$ using the eigenvalues of the
boundary parafermion charge operators
\begin{subequations}
\label{eq:parafermion charge mu even}
\begin{align}
\label{eq:parafermion charge mu even a}
Q^{\,}_{\mathrm{B},a}=
e^{\mathrm{i}\pi\frac{N+1}{N}}\,
\gamma^{\,}_{\mathrm{B},\, 2a-1}\,
\gamma^{\dag}_{\mathrm{B},\, 2a},
\end{align}
with $a=1,\cdots,n^{\,}_{\mathrm{B}}/2$. 
The boundary parafermion charge symmetry
is then implemented by the unitary operator
\begin{align}
\label{eq:parafermion charge mu even b}
Q^{\,}_{\mathrm{B}}= 
\prod^{n^{\,}_{\mathrm{B}}/2}_{a=1}
Q^{\,}_{\mathrm{B},a}.
\end{align}
\end{subequations}
For a general element $\tilde{g}\equiv (z,g)$ in $\mathcal{G}$, 
we parameterize its boundary representation as 
\begin{align}
\label{eq:general proj rep mu even}
U^{\,}_{\mathrm{B}}(z,g)=
Q^{z}_{\mathrm{B}}\,
C^{\,}_{\mathrm{B}}(g)\,
V^{\,}_{\mathrm{B}}(g).
\end{align}
Here, $C^{\,}_{\mathrm{B}}(g)$ is the unitary 
operator that implements the (special) charge conjugation
symmetry if the group action $\rho_{g}\neq 1$ for 
the element $g\in G^{\,}_{b}$. The unitary operator
$V^{\,}_{\mathrm{B}}(g)$
carries a definite parafermion charge, i.e.,
\begin{align}
\label{eq:index kappa mu even}
V^{\,}_{\mathrm{B}}(g)\,
Q^{\,}_{\mathrm{B}}\,=
\omega^{\kappa(g)}_{N}\,
Q^{\,}_{\mathrm{B}}\,
V^{\,}_{\mathrm{B}}(g),
\end{align}
with the 1-cochain $\kappa\in C^{1}(G^{\,}_{b},\,\Z^{\,}_{N})$  
encoding the parafermion 
charge of $V^{\,}_{\mathrm{B}}(g)$.

The boundary representation $U^{\,}_{\mathrm{B}}$ 
is projective in that it obeys the composition rule
\begin{align}
\label{eq:proj alg mu even}
U^{\,}_{\mathrm{B}}(\tilde{g})\,
U^{\,}_{\mathrm{B}}(\tilde{h})=
e^{\mathrm{i}\phi(\tilde{g},\,\tilde{h})}\,
U^{\,}_{\mathrm{B}}(\tilde{g}\circ\tilde{h}).
\end{align}
The projective phase $\phi\in Z^{2}(\mathcal{G}, U(1))$ is a
representative 2-cocycle of the second group cohomology equivalence class
$[\phi]\in H^{2}(\mathcal{G}, U(1))$, see Appendix
\ref{appsec:Group Cohomology} for a review.
Thus, the cohomology group
$H^{2}(\mathcal{G}, U(1))$ classifies the distinct projective
representations for an even number of boundary parafermions.

One can decompose the 2-cocycle
$\phi$ in terms of a tuple
$(\kappa,\nu)\in C^{1}(G^{\,}_{b},\Z^{\,}_{N})\times C^{2}(G^{\,}_{b},U(1))$
as follows. 
Recall that $\kappa$ was already introduced in Eq.\ 
(\ref{eq:index kappa mu even}).
We define $\nu$ as the restriction of $\phi$ on the bosonic symmetries $G^{\,}_{b}\cong \mathcal{G}/\Z^{\mathrm{pf}}_{N}$, i.e.,
\begin{align}
\label{eq:index nu mu even}
U^{\,}_{\mathrm{B}}(0,g)\,
U^{\,}_{\mathrm{B}}(0,h)=
e^{\mathrm{i}\nu(g,h)}\,
U^{\,}_{\mathrm{B}}(c(g,h),\,g\,h).
\end{align}
Inserting the parameterization \eqref{eq:general proj rep mu even}
into Eq.\ \eqref{eq:proj alg mu even} yields
\begin{subequations}
\label{eq:phi in terms of nu kappa mu even}
\begin{align}
\label{eq:phi in terms of nu kappa mu even a}
\begin{split}
U^{\,}_{\mathrm{B}}(\tilde{g})\,
U^{\,}_{\mathrm{B}}(\tilde{h})=
\omega^{\kappa(g)\,u}_{N}\,
e^{\mathrm{i}\nu(g,\,h)}\,
U^{\,}_{\mathrm{B}}(\tilde{g}\circ \tilde{h}).
\end{split}
\end{align}
Comparison with Eq.\ \eqref{eq:proj alg mu even}
produces the identity
\begin{align}
\label{eq:phi in terms of nu kappa mu even b}
\phi(\tilde{g},\tilde{h})
=
\nu(g, h)
+
\frac{2\pi}{N}\,
\kappa(g)\,
u.
\end{align}
\end{subequations}
Thus, the tuple $(\kappa,\,\nu)$ carries the same information as 
the projective phase $\phi$. However, while $\phi$ is a 2-cocycle, 
$\nu$ does not have a vanishing coboundary. Instead, the pair $(\kappa,\nu)$
satisfies a \textit{modified} 2-coboundary condition
\begin{subequations}
\begin{align}
\label{eq:modified 2-coboundary}
\begin{split}
&
\mathcal{D}^{(2)}\,(\kappa,\nu)=0,
\\
&
\mathcal{D}^{(2)}\,(\kappa,\nu):= 
(\delta^{(1)}_\rho\,\kappa,\,\delta^{(2)}\nu-\frac{2\pi}{N}\,\kappa\cup c).
\end{split}
\end{align}
Similarly, the tuple $(\kappa,\,\nu)$
is equivalent to 
another tuple $(\kappa',\,\nu')$
up to
a \textit{modified} 1-coboundary
\begin{align}
\label{eq:modified 1-coboundary}
\begin{split}
&
(\kappa',\,\nu')
=
(\kappa,\,\nu) + \mathcal{D}^{(1)}(\beta,\, \alpha),
\\
&
\mathcal{D}^{(1)}(\beta,\, \alpha):=
(0,\delta^{(1)}\alpha -  \frac{2\pi}{N} \beta\cup c),
\end{split}
\end{align}
where $\alpha\in C^{1}(G^{\,}_{b},U(1))$ and
$\beta\in C^0(G^{\,}_{b}, \Z^{\,}_{N})$.
One verifies that
$\mathcal{D}^{(2)}\circ\mathcal{D}^{(1)}=0$, 
and, thus, in terms of the pair $(\kappa,\nu)$,
the second cohomology group
$H^{2}(\mathcal{G},U(1))$
can be recast as
\begin{align}
H^{2}(\mathcal{G}, U(1))
\cong
\frac{\mathrm{ker} \, \mathcal{D}^{(2)}}{\mathrm{im} \, \mathcal{D}^{(1)}}.
\end{align}
\end{subequations}
The index $\nu$ in the decomposition
\eqref{eq:phi in terms of nu kappa mu even}
encodes how the bosonic symmetry $G^{\,}_{b}$ is realized projectively, 
while the index $\kappa$ encodes the parafermion charge carried by 
the representations of $G^{\,}_{b}$. 
The triplet of indices
\begin{align}
(\mu,\, [(\kappa,\nu)]) = (0,\, [(\kappa,\nu)]),
\label{eq:triplet-projective-indices-even-case}
\end{align}
thus characterizes the boundary projective representations given an
even number of boundary parafermions.

\subsection{Odd number of boundary parafermions}
\label{subsec:defs indices odd}

When the number of boundary parafermion operators is odd 
($\mu=1$), one cannot define a local Hilbert
space using only the operators
from the boundary set $\mathcal{O}^{\,}_{\mathrm{B}}$.
For example, trying to a construct parafermion charge operators
as in Eq.\ \eqref{eq:parafermion charge mu even a}
will always leave one parafermion operator unpaired.

To define a Hilbert space on which the boundary parafermion
operators act, we add an \textit{auxiliary} parafermion operator 
$\gamma^{\,}_{\infty}$ to the set $\mathcal{O}^{\,}_{\mathrm{B}}$
of boundary parafermion operators. We assume, without loss of generality, 
that this auxiliary parafermion operator satisfies the algebra
\begin{subequations}
\begin{align}
\gamma^{\,}_{\mathrm{B},j}\,
\gamma^{\,}_{\infty}=
\w\,
\gamma^{\,}_{\infty}\,
\gamma^{\,}_{\mathrm{B},j}, 
\quad j=1,\cdots, n_{\mathsf{B}},
\end{align}
and define the extended set 
\begin{align}
\label{eq:parafermion ops ext mu odd}
\mathcal{O}^{\mathrm{ext}}_{\mathrm{B}}=
\left\{
\gamma^{\,}_{\mathrm{B},\,1}, \, 
\gamma^{\,}_{\mathrm{B},\,2},
\cdots,\,
\gamma^{\,}_{\mathrm{B},\,n^{\,}_{\mathrm{B}}},\,
\gamma^{\,}_{\infty}
\right\}.
\end{align}	
\end{subequations}
Physically, such an auxiliary parafermion is provided by the 
unpaired boundary parafermion operator on the opposite boundary,
since by Eq.\ \eqref{eq:parafermion number consistency equation}
both boundaries support an odd number of  
parafermion degrees of freedom. 
The parafermion algebras 
$\mathfrak{P}^{n^{\,}_{\mathrm{B}}}_{N}$
and
$\mathfrak{P}^{n^{\,}_{\mathrm{B}}+1}_{N}$
are generated by the set $\mathcal{O}^{\,}_{\mathrm{B}}$
and $\mathcal{O}^{\mathrm{ext}}_{\mathrm{B}}$, respectively.

Using the extended set of parafermion operators 
\eqref{eq:parafermion ops ext mu odd}, we define 
$(n^{\,}_{\mathrm{B}}+1)/2$ boundary parafermion charge operators
\begin{subequations}
\label{eq:parafermion charge mu odd}
\begin{align}
\begin{split}
&
Q^{\,}_{\mathrm{B},a}=
e^{\mathrm{i}\pi\frac{N+1}{N}}\,
\gamma^{\,}_{\mathrm{B},\, 2a-1}\,
\gamma^{\dag}_{\mathrm{B},\, 2a},
\\
&
Q^{\,}_{\mathrm{B},(n^{\,}_{\mathrm{B}}+1)/2}=
e^{\mathrm{i}\pi\frac{N+1}{N}}\,
\gamma^{\,}_{\mathrm{B},\, n^{\,}_{\mathrm{B}}}\,
\gamma^{\dag}_{\infty},
\end{split}
\end{align}
with $a=1,\,2,\,\cdots,\,(n^{\,}_{\mathrm{B}}-1)/2$. 
We can then label the states in the Hilbert space 
$\mathcal{H}^{\,}_{\mathrm{B}}$ using the eigenvalues of these operators.
Unlike the $\mu=0$ case, $\mathcal{H}^{\,}_{\mathrm{B}}$ 
is non-local as some of the operators acting on it 
are supported on both boundaries of an open chain.
The boundary parafermion charge symmetry is then 
implemented by the unitary operator
\begin{align}
Q^{\,}_{\mathrm{B}}= 
\prod^{(n^{\,}_{\mathrm{B}}+1)/2}_{a=1}
Q^{\,}_{\mathrm{B},a}.
\end{align}
\end{subequations}

When constructing the boundary representations for a general
element $(z,g)\in \mathcal{G}$, we demand that the key relation
\eqref{eq:bd rep key relation} applies only to the parafermions 
from the set $\mathcal{O}^{\,}_{\mathrm{B}}$
and not to the auxiliary parafermion $\gamma^{\,}_{\infty}$. 
Since this is a less restrictive requirement than demanding 
that Eq.~\eqref{eq:bd rep key relation} applies to the extended
set \eqref{eq:parafermion ops ext mu odd}, there is an additional
redundancy when constructing the boundary representations
in comparison to the ${\mu=0}$ case. This redundancy is rooted in the 
fact that the algebra $\mathfrak{P}^{n^{\,}_{\mathrm{B}}}$ has 
a non-trivial center when $n^{\,}_{\mathrm{B}}$ is odd. This
$N$-dimensional center is generated by the operator 
\begin{align}
\label{eq:center Y mu odd}
Y=
\left(
\prod_{a=1}^{(n^{\,}_{\mathrm{B}}-1)/2}
Q^{\,}_{\mathrm{B},a}
\right)
\gamma^{\,}_{n^{\,}_{\mathrm{B}}},
\end{align}
which commutes with all the boundary parafermion operators
in the set $\mathcal{O}^{\,}_{\mathrm{B}}$. Note that the operator
$Y$
is a product of all boundary parafermion operators and 
does not contain the auxiliary parafermion
$\gamma^{\,}_{\infty}$.
It carries unit parafermion charge, 
\begin{align}
\label{eq:parafermion charge of Y}
Q^{\,}_{\mathrm{B}}\,
Y\,
Q^{\dag}_{\mathrm{B}}=
\omega^{\,}_{N}\,
Y.
\end{align}
Since $Y$ commutes with all boundary parafermion operators 
from the set $\mathcal{O}^{\,}_{\mathrm{B}}$, for any boundary representation
$U^{\,}_{\mathrm{B}}(\tilde{g})$ satisfying Eq.\
\eqref{eq:bd rep key relation}, the operators
\begin{align}
\label{eq:redefinition by central element}
U^{\prime}_{\mathrm{B}}(\tilde{g}):= 
Y\,U^{\,}_{\mathrm{B}}(\tilde{g})
\end{align}
satisfy the same equation. In other words, 
$U^{\prime}_{\mathrm{B}}(\tilde{g})$ is also a
valid boundary representation of $\mathcal{G}$. However, 
$U^{\prime}_{\mathrm{B}}(\tilde{g}) $ and 
$U^{\,}_{\mathrm{B}}(\tilde{g})$ differ in their 
algebra with the parafermion
charge operator $Q^{\,}_{\mathrm{B}}$.

For a general element $\tilde{g}\equiv (z,g)$ in $\mathcal{G}$,
we parameterize its boundary representation as
\begin{align}
\label{eq:general proj rep mu odd}
U^{\,}_{\mathrm{B}}(z,g)=
Q^{z}_{\mathrm{B}}\,
C^{\,}_{\mathrm{B}}(g)\,
V^{\,}_{\mathrm{B}}(g)\,
\gamma^{-\kappa(g)}_{\infty},
\end{align}
see Appendix \ref{appsec:odd parafermion rep identities} for a derivation.
Here, $C^{\,}_{\mathrm{B}}(g)$ is the unitary 
operator that implements the (special) charge conjugation
symmetry if the group action $\rho_{g}\neq 1$ for 
the element $g\in G^{\,}_{b}$. The exponent $\kappa(g)$
is a $1$-cochain $\kappa\in C^{1}(G^{\,}_{b},\Z^{\,}_{N})$,
the interpretation of which will be given below.
The unitary $V^{\,}_{\mathrm{B}}(g)$
commutes with both $Y$ and $\gamma^{\,}_{\infty}$
such that the boundary representation satisfies
\begin{align}
\label{eq:mu odd algebra with parafermion charge}
U^{\,}_{\mathrm{B}}(z,g)\,
Q^{\,}_{\mathrm{B}}=
\omega^{\kappa(g)}\,
Q^{\rho_{g}}_{\mathrm{B}}\,
U^{\,}_{\mathrm{B}}(z,g).
\end{align}
In other words, the unitary operator
$V^{\,}_{\mathrm{B}}(g)\,\gamma^{-\kappa(g)}_{\infty}$
carries parafermion charge $-\kappa(g)$. 
Since the unitary $V^{\,}_{\mathrm{B}}(g)$ commutes with $Y$,
it is an element in $\mathfrak{P}^{n^{\,}_{\mathrm{B}}}_{N}$ while 
$U^{\,}_{\mathrm{B}}(z,g)\in \mathfrak{P}^{n^{\,}_{\mathrm{B}}+1}_{N}$.
The fact that $\gamma^{\,}_{\infty}$ factorizes 
as in Eq.~\eqref{eq:general proj rep mu odd}
stems from the requirement that the algebra $\mathfrak{P}^{n^{\,}_{\mathrm{B}}}_{N}$
must be closed under the boundary representation 
$U^{\,}_{\mathrm{B}}(z,g)$
[recall Eq.\ \eqref{eq:bd compatibility cond}].

The presence of the non-trivial center puts restrictions on 
which parafermionic symmetries $\mathcal{G}$ are compatible with
an odd number of parafermions. Conjugation of both sides of
\begin{subequations}
\begin{align}
U^{\,}_{\mathrm{B}}(0,g)\,
U^{\,}_{\mathrm{B}}(0,h)=
e^{\mathrm{i}\phi(\tilde{g}, \tilde{h})}\,
Q^{c(g,h)}_{\mathrm{B}}\,
U^{\,}_{\mathrm{B}}(0,\,g\,h)
\end{align}
by $Y$
implies that
\begin{equation}
c=\delta^{(1)}_\rho\,\kappa=0.
\end{equation}
Hence, the only symmetry groups
that are compatible with $\mu=1$ are those that split, i.e., of the form
\begin{equation}
\mathcal{G}\cong Z^{\mathrm{pf}}_{N}\rtimes G^{\,}_{b}.
\end{equation}
\end{subequations}
Conversely, when $\mathcal{G}$ 
is a nonsplit extension,
only boundaries with an even number of parafermions (${\mu=0}$)
are possible.

The composition of two elements $\tilde{g}, \tilde{h}\in \mathcal{G}$
follows Eq.~\eqref{eq:phi in terms of nu kappa mu even a}
just as in $\mu=0$ case. Thus, the projective phase in the group composition
\begin{subequations}
\begin{align}
\label{eq:proj alg mu odd}
U^{\,}_{\mathrm{B}}(\tilde{g})\,
U^{\,}_{\mathrm{B}}(\tilde{h})=
e^{\mathrm{i}\phi(\tilde{g},\,\tilde{h})}\,
U^{\,}_{\mathrm{B}}(\tilde{g}\circ\tilde{h})
\end{align}
has the same decomposition
\begin{align}
\phi(\tilde{g}, \tilde{h}) = 
\nu(g,h)
+
\frac{2\pi}{N}\,\kappa(g)\,u.
\end{align}
\end{subequations}
However, because the group $\mathcal{G}$ splits $(c=0)$, 
Eq.\ \eqref{eq:modified 2-coboundary} implies that 
both $\nu$ and $\kappa$ are proper cocycles. 
The index $\nu(g,h)$ can be computed by restricting $\phi$ to
the bosonic symmetries $G^{\,}_{b}$,
\begin{align}
\label{eq:index nu mu odd}
U^{\,}_{\mathrm{B}}(0,g)\,
U^{\,}_{\mathrm{B}}(0,h)\,=
e^{\mathrm{i}\nu(g,h)}\,
U^{\,}_{\mathrm{B}}(0, g\,h).
\end{align}
While the index $\kappa(g)$ encodes the parafermion charge of the
product $V^{\,}_{\mathrm{B}}(g)\,\gamma^{-\kappa(g)}_{\infty}$
in the parameterization \eqref{eq:general proj rep mu odd}, 
this definition is not invariant under the redefinition of
$U^{\,}_{\mathrm{B}}(z,g)$ 
in Eq.\ \eqref{eq:redefinition by central element}.
We thus define the index $\kappa(g)$ through the relation
\begin{align}
\label{eq:index kappa mu odd}
U^{\,}_{\mathrm{B}}(0,g)\,
Y=
\omega^{\kappa(g)}_{N}\,
Y^{\rho_{g}}\,
U^{\,}_{\mathrm{B}}(0,g),
\end{align}
which is invariant under the redefinition
\eqref{eq:redefinition by central element}.

To summarize, when $\mu=1$, the group must split, i.e.,
\begin{subequations}
\begin{equation}
\mathcal{G}\cong \Z^{\mathrm{pf}}_{N}\rtimes G^{\,}_{b},
\end{equation}
and the triplet of indices
\begin{align}
(\mu,\,[(\kappa,\nu)])=(1,\,[\kappa],[\nu]),
\label{eq:triplet-projective-indices-odd-case}
\end{align}
\end{subequations}
characterizes the boundary projective representations of
given an odd number of boundary parafermions.

\subsection{Ground state degeneracies}
\label{subsec:Groundstate degeneracies}

We can relate the triplet of indices $(\mu, [(\kappa,\nu)])$
to (protected) ground state degeneracy (GSD) of an IPT 
phase when open boundary conditions are imposed.
When the indices $(\mu, [(\kappa,\nu)])$
are trivial, i.e.,
\begin{align}
\label{eq:indices are trivial}
(\mu, [(\kappa,\nu)]) = (0, [(0,0)]),
\end{align}
there is no constraint on the GSD associated with the boundary modes.
In other words, a singlet representation of $\mathcal{G}$
is compatible with the trivial indices. We associate this case 
with the trivial IPT phase. 
When Eq.\ \eqref{eq:indices are trivial} does not hold,
there is a minimal GSD larger than one that cannot be lifted 
as we shall describe.

When open boundary conditions are imposed, there are two sets
of indices $(\mu_\text{L}, [(\kappa_\text{L},\nu_\text{L})])$ 
and $(\mu_\text{R}, [(\kappa_\text{R},\nu_\text{R})])$
associated with the left and the right boundaries, respectively.
We distinguish two cases, $\mu_\text{L}=\mu_\text{R}=0,1$
that corresponds to having even and odd number of parafermionic degrees 
of freedom at the two boundaries 
(recall Eq.\ \eqref{eq:parafermion number consistency equation}), respectively.

When ${\mu_\text{L}=\mu_\text{R}=0}$, there are well-defined local 
Hilbert spaces associated with the two ends. In this case, 
the total Hilbert space associated with the boundaries 
factorizes
\begin{align}
\mathcal{H}_\text{bd}
=
\mathcal{H}_\text{L}
\otimes
\mathcal{H}_\text{R}.
\end{align}
We can consider the boundary Hamiltonians 
$H_\text{L}\in \mathfrak{P}^{n_\text{L}}_N$ and 
$H_\text{R}\in \mathfrak{P}^{n_\text{R}}_N$ that act on 
the Hilbert spaces $\mathcal{H}_\text{L}$ and 
$\mathcal{H}_\text{R}$, respectively. By definition, 
the Hamiltonians commute with the boundary projective representations
$U^{\,}_\text{L}(z,g)$ and $U^{\,}_\text{R}(z,g)$, respectively.
The minimal global GSD is then the minimal GSD of 
the Hamiltonian
\begin{align}
H_\text{bd} 
= 
H_\text{L} \otimes \mathbb{1}_{\mathcal{H}_\text{R}}
+
\mathbb{1}_{\mathcal{H}_\text{L}} \otimes H_\text{R}.
\end{align}
The minimal degeneracy associated with the left and the right boundaries
then can be deduced by computing the minimum dimensional projective 
irreducible representation (irrep) that is compatible with the indices 
$[(\kappa_\text{L}, \nu_\text{L})]$ and $[(\kappa_\text{R}, \nu_\text{R})]$.
Let
$d_{[(\kappa_\text{L}, \nu_\text{L})]}$ 
and 
$d_{[(\kappa_\text{R}, \nu_\text{R})]}$
be the smallest dimensions of the irreps
associated with
the left and the right boundary, respectively.
We then have global GSD given by
\begin{align}
\text{GSD}^{\mu=0}_\text{bd}
=
d_{[(\kappa_\text{L}, \nu_\text{L})]}
\times
d_{[(\kappa_\text{R}, \nu_\text{R})]}.
\end{align}
For a given boundary $\text{B}=\text{L},\text{R}$ 
the dimension $d_{[(\kappa_\text{B}, \nu_\text{B})]}$
depends on the details of the group $\mathcal{G}$ and the 
indices $[(\kappa_\text{B}, \nu_\text{B})]$.

When ${\mu_\text{L}=\mu_\text{R}=1}$, the Hilbert space associated
with the boundaries are necessarily non-local. The total
Hilbert space associated with the boundaries then factorizes as
\begin{align}
\mathcal{H}_\text{bd}
=
\mathcal{H}_\text{L}
\otimes
\mathcal{H}_\text{LR}
\otimes
\mathcal{H}_\text{R},
\end{align}
where $\mathcal{H}_\text{L}$ and $\mathcal{H}_\text{R}$ are spanned
by the eigenstates of the parafermion charge operators
localized at the left and right boundaries respectively. 
The states at the non-local Hilbert space $\mathcal{H}_\text{LR}$ 
are spanned by the eigenstates of the single parafermion charge operator
constructed from the unpaired parafermion operators in 
$\mathfrak{P}^{n_\text{L}}_N$ and  $\mathfrak{P}^{n_\text{R}}_N$.
The boundary Hamiltonians 
$H_\text{L}\in \mathfrak{P}^{n_\text{L}}_N$ and 
$H_\text{R}\in \mathfrak{P}^{n_\text{R}}_N$
act on the Hilbert spaces
\begin{subequations}
\begin{align}
\mathcal{H}_\text{L}
\otimes
\mathcal{H}_\text{LR},
\end{align}
and 
\begin{align}
\mathcal{H}_\text{R}
\otimes
\mathcal{H}_\text{LR},
\end{align}
\end{subequations}
respectively. Since ${\mu_\text{L}=\mu_\text{R}=1}$,
there exists operators $Y_\text{L}$ and $Y_\text{R}$ that 
commute with the Hamiltonians $H_\text{L}$ and $H_\text{R}$,
respectively. 
However, the central elements do not commute with each other as well
as the global parafermion charge operator. In particular,
we have
\begin{align}
\label{eq:algebra between two central elements and Q}
\begin{split}
&
Y_\text{L}\,Y_\text{R} = \omega_N Y_\text{R}\, Y_\text{L}
\\
&
Q\,Y_\text{B} = \omega_N\, Y_\text{B}\,Q,
\end{split}
\end{align}
where $Q$ is the global parafermion charge operator. 
Because of this global algebra, $\mathcal{H}_\text{bd}$ 
must be at least $N$-fold degenerate, which is associated with
the non-trivial indices ${\mu_\text{L}=\mu_\text{R}=1}$.
Importantly, this $N$-fold degeneracy can be lifted by explicitly
breaking the $\Z^{\mathrm{pf}}_N$ symmetry (recall that $N=p^k$).
However, a $p$-fold degeneracy
cannot be lifted by any local perturbation since $\Z^{\mathrm{pf}}_N$
can be broken down at most down to $\Z^{\mathrm{pf}}_p$ symmetry 
by a local term.

There can be additional degeneracies protected by the non-trivial 
values of the indices $[(\kappa_\text{B}, \nu_\text{B})]$. 
As in the case of ${\mu_\text{L}=\mu_\text{R}=0}$, we associate
denote by $d_{[(\kappa_\text{B}, \nu_\text{B})]}$
the smallest dimension of an irrep that is compatible with 
the indices $[(\kappa_\text{B}, \nu_\text{B})]$
at boundary $\text{B}$ can have. 
The global GSD is then given by
\begin{align}
\text{GSD}^{\mu=1}_\text{bd}
=
\frac{1}{N}
\times
d_{[(\kappa_\text{L}, \nu_\text{L})]}
\times
d_{[(\kappa_\text{R}, \nu_\text{R})]},
\end{align}
where we divide by $N$ to account for double counting 
the global degeneracy associated with the non-local Hilbert space 
$\mathcal{H}_{\text{LR}}$. 

\subsection{Classification for general $N$}
\label{subsec:general N}

We have so far considered parafermions of prime-power order, 
i.e., ${N= p^k}$, where $p$ 
is a prime and $k$ is a positive integer. 
We seek to generalize the triplet $(\mu, (\kappa,\nu))$ to any finite $N$.
To this end, consider the prime-power decomposition
\begin{subequations}
\label{eq:prime decomposition of N}
\begin{align}
\label{eq:prime decomposition of N a}
N = p^{k_1}_1\,p^{k_2}_2\,\cdots\,p^{k_n}_n.
\end{align}
The global parafermion charge symmetry factorizes accordingly,
\begin{align}
\label{eq:prime decomposition of N b}
\Z^{\mathrm{pf}}_N
=
\Z^{\mathrm{pf}}_{N_1}
\times
\Z^{\mathrm{pf}}_{N_2}
\times
\cdots
\times
\Z^{\mathrm{pf}}_{N_n},
\end{align}
\end{subequations}
where $N_a = p^{k_a}_a$ for $a=1,\cdots,n$.
Let us consider the operators 
\begin{align}
\label{eq:lower order parafermion operators}
\gamma^{(a)}_j
:=
\gamma^{N/N_a}_j
\end{align}
for all $j=1,\dots, 2L$. Note that each operator
\eqref{eq:lower order parafermion operators} is of order $N_a$
and has the parafermion charge
\begin{align}
\label{eq:lower order parafermion operators charge}
\begin{split}
Q\,
\gamma^{(a)}_j\,
Q^\dagger
&=
\omega^{N/N_a}_N\,
\gamma^{(a)}_j\\
&=
\omega^{\,}_{N_a}\,
\gamma^{(a)}_j.
\end{split}
\end{align}
These operators satisfy the algebra
\begin{align}
\label{eq:gen N flavor algebra}
\begin{split}
\gamma^{(a)}_j\,
\gamma^{(b)}_k
&=
e^{\mathrm{i}\frac{2\pi}{N} \frac{N^2}{N_a\,N_b} \text{sgn}(k-j)}
\gamma^{(b)}_k\,
\gamma^{(a)}_j
\\
&=
e^{\mathrm{i}\frac{2\pi}{N_a} \frac{N}{N_b} \text{sgn}(k-j)}
\gamma^{(b)}_k\,
\gamma^{(a)}_j.
\end{split}
\end{align}
Due to the prime decomposition
\eqref{eq:prime decomposition of N a}, we have 
\begin{align}
\frac{N}{N_b} = s_a\,\delta_{a,b} \mod N_a,
\end{align}
where $s_a=N/N_a\mod N_a$ is an integer coprime to $N_a$. 
We note that when $a\neq b$, the set of operators
$\gamma^{(a)}_j$ and $\gamma^{(b)}_k$ commute for any $j$ and $k$. When $a=b$, the algebra \eqref{eq:gen N flavor algebra}
reduces to 
\begin{align}
\label{eq:gen N algebra for a single flavor}
\gamma^{(a)}_j\,
\gamma^{(a)}_k
=
\omega^{s_a\,\mathrm{sgn}(k-j)}_{N_a}\,
\gamma^{(a)}_k\,
\gamma^{(a)}_j.
\end{align}
While this is reminiscent of the parafermion algebra
\eqref{eq:Order N Pf Algebra}, the phase factor on
the right-hand side is not the principal $N_a$th 
root of unity, i.e., $\omega_{N_a}$
(it is, however, still a primitive root).
Interestingly, the algebra 
\eqref{eq:gen N algebra for a single flavor}
is in general distinct from the parafermion algebra 
\eqref{eq:Order N Pf Algebra} with the 
principal $N_a$th root of unity in the sense
that they are not related by a local redefinition of 
the parafermion operators~\footnote{
The simplest example to see this is $\Z_3$ parafermions. 
Consider the two algebras 
$\gamma^{\,}_j\,\gamma^{\,}_k = 
\omega^{\mathrm{sgn}(k-j)}_3\,
\gamma^{\,}_k\,\gamma^{\,}_j$
and 
$\tilde{\gamma}^{\,}_j\,\tilde{\gamma}^{\,}_k = 
\omega^{-\mathrm{sgn}(k-j)}_3\,
\tilde{\gamma}^{\,}_k\,\tilde{\gamma}^{\,}_j$.
There is an isomorphism
between the two algebras, but it must necessarily 
reverse the order of the parafermions, i.e., 
it must act as a spatial reflection. 
In particular, there is no
local redefinition
$\tilde{\gamma}^{\,}_j\equiv\gamma^{\alpha_j}_j$
that takes one algebra to the other.
}. 

We interpret for each $a =1,\cdots,n$ the set of operators
$\{\gamma^{(a)}_j\}_{j=1}^{2L}$ from Eq.\ 
\eqref{eq:lower order parafermion operators}
as a decoupled flavor of order-$N_a$ parafermions.
In particular, the total
Hilbert space factorizes into subspaces,
\begin{subequations}
\begin{align}
\mathcal{H} = \bigotimes_{a=1}^n
\mathcal{H}_a,
\end{align}
where each factor $\mathcal{H}_a$
is spanned by the eigenstates of the 
order-$N_a$ parafermion
charge operator~\footnote{
Recall Eq.\ \eqref{eq:Pf local charge a}. 
Here, we use an analogous notation for
each flavor, i.e., $\xi^{(a)}_j\equiv \gamma^{(a)}_{2j-1}$
and $\eta^{(a)}_j\equiv \gamma^{(a)}_{2j}$.
}
\begin{align}
Q^{(a)}
=
\prod_{j=1}^{L}
e^{\mathrm{i}\pi\,s_a\,(N_a+1)/N_a}\,
\xi^{(a)}_j\,
(\eta^{(a)}_j)^\dag.
\end{align}
We note that the action of the operator 
$Q^{(a)}$ on the parafermion operators
$\gamma^{(a)}_j$ do not match the action 
\eqref{eq:lower order parafermion operators charge}
of the operator $Q$
since 
\begin{align}
Q^{(a)}\,
\gamma^{(a)}_j\,
\left(
Q^{(a)}
\right)^\dag
=
\omega^{s_a}_{N_a}\,
\gamma^{(a)}_j.
\end{align}
\end{subequations}
Nevertheless, there exists
a factorization
\begin{align}
\label{eq:order N pf charge decomposition to lower}
Q
=
e^{\mathrm{i}\varphi}\,
\prod_{a=1}^n
\left(
Q^{(a)}
\right)^{b_a},
\end{align}
with integers ${b_a=1,\cdots, N_a-1}$.
This is guaranteed by the fact that $s_a$ and $N_a$ are coprime,
i.e., there exist integers $b_a$ such that 
${b_a\,s_a = 1 \! \mod N_a}$ for all $a=1,\cdots,n$. 
Since the left-hand and right-hand sides of Eq.\ 
\eqref{eq:order N pf charge decomposition to lower} 
have identical action for all flavors $a$, they must be equal up to a 
$U(1)$ phase which we denote by and $e^{\mathrm{i}\varphi}$.

For symmetry operators other than $Q$, a 
factorization similar to the one
in Eq.\ \eqref{eq:order N pf charge decomposition to lower}
is not guaranteed. The factorization 
\eqref{eq:order N pf charge decomposition to lower} 
means that the parafermion charge of each flavor is separately conserved.
However, a symmetry operator can still mix two flavors $a$ and $b$ 
provided
that the parafermion operator $\gamma^{(a)}_j$
is mixed with operators from flavor $b$ with zero
$\Z^{\mathrm{pf}}_{N_b}$ parafermion charge.

Nevertheless, the classification of IPT phases for 
general $N$ simplifies when a factorization similar
to Eq.\ \eqref{eq:order N pf charge decomposition to lower}
holds for any symmetry operators~\footnote{
We leave the case when the factorization
\eqref{eq:factorization holds for any symmetry}
does not hold for future work.
}, i.e.,
\begin{align}
\label{eq:factorization holds for any symmetry}
U(\tilde{g})
=
e^{\mathrm{i}\varphi}\,
\prod_{a=1}^n
U^{(a)}(\tilde{g}).
\end{align}
In this case, the unitary representations $U^{(a)}(\tilde{g})$
in general form a subgroup of the full symmetry group $\mathcal{G}$,
i.e., $\mathcal{G}$ in general acts as a subgroup $\mathcal{G}^{(a)}$
on the flavor $a$. 
For example, this is true for $\Z^{\mathrm{pf}}_N$ parafermion charge
symmetry which acts as the $\Z^{\mathrm{pf}}_{N_a}$ subgroup.

When the factorization \eqref{eq:factorization holds for any symmetry}
holds, the classification of IPT phases for 
order-$N$ parafermions thus reduces to the 
classification of IPT phases for each order-$N_a$
parafermion flavor with $\mathcal{G}^{(a)}\subset \mathcal{G}$
symmetry, which has been achieved in Sec.\ 
\ref{sec:Boundary projective representations}.
We introduce a triplet of indices
$\big(\mu^{(a)}, [\kappa^{(a)},\nu^{(a)}]\big)$ 
such that
\begin{subequations}
\label{eq:general N collection of indices}
\begin{align}
\begin{split}
&\mu^{(a)}=0,1,
\\
&\kappa^{(a)} \in C^{1}(\mathcal{G}^{(a)}/\Z^{\mathrm{pf}}_{N_a},\,\Z_{N_a}),
\\
&\nu^{(a)} \in
C^{2}(\mathcal{G}^{(a)}/\Z^{\mathrm{pf}}_{N_a},\, U(1)),    
\end{split}
\end{align}
that can take values independently for each flavor.
The complete classification for order-$N$ parafermions
is then given by the collection of triplets
\begin{align}
\left\{
\big(\mu^{(a)}, [\kappa^{(a)},\nu^{(a)}]\big)
\right\}.
\end{align}
\end{subequations}
We note that when only the $\Z^{\mathrm{pf}}_N$ 
parafermion charge symmetry is present, only  the indices 
$\mu^{(a)}=0,1$ in the decomposition of 
indices \eqref{eq:general N collection of indices} for each flavor
can be non-trivial. We find that when no additional bosonic symmetries 
are present there is 
a non-trivial IPT phase for each prime factor in $N$,
with the classification given by $\Z^n_2$.

\begin{figure*}[t]
\centering

\centering
\resizebox{\linewidth}{!}{%
\begin{tikzpicture}[
x=1cm,
y=1cm,
every node/.style={font=\large},
site/.style={
circle,
inner sep=0pt,
outer sep=0pt,
minimum size=6pt
},
large site/.style={
site,
minimum size=14pt
},
bond/.style={
draw=black!85,
line width=0.9pt,
line cap=round,
shorten <=3pt,
shorten >=3pt
},
highlight/.style={
draw=ringred,
line width=1.2pt
}
]


\node at (0,0) {$\mathrm{a)}$};

\begin{scope}[xshift=1.35cm]

\node[large site,fill=siteorange] (a1) at (1.7, 0.4) {};
\node[large site,fill=siteteal]   (a2) at (2.5,-0.4) {};
\node[site,fill=siteorange]      (a3) at (3.3, 0.4) {};
\node[site,fill=siteteal]        (a4) at (4.1,-0.4) {};
\node[site,fill=siteorange]      (a5) at (4.9, 0.4) {};
\node[site,fill=siteteal]        (a6) at (5.7,-0.4) {};

\foreach \u/\v in {a1/a2,a2/a3,a3/a4,a4/a5,a5/a6}{
	\draw[bond] (\u) -- (\v);
}

\draw[highlight] (a1) circle[radius=9.5pt];
\draw[highlight] (a2) circle[radius=9.5pt];

\node[text=siteorange] at (1.40,1.12) {$\Phi_1^{\text{L}}$};
\node[text=siteteal]   at (1.55,-0.75) {$\Phi_2^{\text{L}}$};

\node at (6.45,0) {$\cdots$};
\node[font=\Huge] at (7.65,0) {$\cong$};

\node[large site,fill=sitepurple] (ap1) at (9.00,0) {};
\node[site,fill=sitepurple]       (ap2) at (10.2,0) {};
\node[site,fill=sitepurple]       (ap3) at (11.1,0) {};
\node[site,fill=sitepurple]       (ap4) at (12.0,0) {};

\foreach \u/\v in {ap1/ap2,ap2/ap3,ap3/ap4}{
	\draw[bond] (\u) -- (\v);
}

\draw[highlight] (ap1) circle[radius=9.5pt];

\node[text=sitepurple] at (8.85,0.75)
{$\Phi_{1\wedge 2}^{\text{L}}$};

\node at (12.65,0) {$\cdots$};
\end{scope}


\begin{scope}[shift={(0,-2.5)}]

\node at (0,0) {$\mathrm{b)}$};

\node at (1.2,0) {$\cdots$};

\node[site,fill=siteorange]       (bo1) at (1.80,0) {};
\node[site,fill=siteorange]       (bo2) at (2.70,0) {};
\node[large site,fill=siteorange] (bo3) at (3.90,0) {};

\node[large site,fill=siteteal]   (bt1) at (5.10,0) {};
\node[site,fill=siteteal]         (bt2) at (6.30,0) {};
\node[site,fill=siteteal]         (bt3) at (7.20,0) {};

\foreach \u/\v in {bo1/bo2,bo2/bo3,bt1/bt2,bt2/bt3}{
	\draw[bond] (\u) -- (\v);
}

\draw[highlight] (bo3) circle[radius=9.5pt];
\draw[highlight] (bt1) circle[radius=9.5pt];

\node[text=siteorange] at (3.90,0.75) {$\Phi_1^{\text{R}}$};
\node[text=siteteal]   at (5.10,0.75) {$\Phi_2^{\text{L}}$};

\node at (7.80,0) {$\cdots$};
\node[font=\Huge] at (9.00,0) {$\cong$};

\node at (10.2,0) {$\cdots$};

\node[site,fill=siteorange]       (br1) at (10.7,0) {};
\node[site,fill=siteorange]       (br2) at (11.6,0) {};
\node[large site,fill=sitepurple] (br3) at (12.8,0) {};
\node[site,fill=siteteal]         (br4) at (14.0,0) {};
\node[site,fill=siteteal]         (br5) at (14.9,0) {};

\foreach \u/\v in {br1/br2,br2/br3,br3/br4,br4/br5}{
	\draw[bond] (\u) -- (\v);
}

\draw[highlight] (br3) circle[radius=9.5pt];

\node[text=sitepurple] at (12.8,0.75)
{$\Phi_{1\wedge 2}^{\text{I}}$};

\node at (15.45,0) {$\cdots$};

\end{scope}

\end{tikzpicture}%
}

\caption{\label{fig:two stacking scenarios}  
Two scenarios for stacking parafermionic chain~1 (orange) with
parafermionic chain~2 (green). Prior to stacking, all parafermionic
boundary states are assumed to be localized near the first and last
sites of the respective chains marked by the red circles. 
(a) Stacking rules at the left
boundary under the convention that all boundary states of chain~1 are
ordered to the left of those of chain~2. The labels
$\Phi^{\text{L}}_1$ and $\Phi^{\text{L}}_2$
denote the left-boundary projective representations of chains~1 and~2, respectively, 
while $\Phi^{\text{L}}_{1\wedge 2}$
denotes their corresponding projective stacked representation.
(b) Chain~1 is placed to the left of chain~2. The labels
$\Phi^{\text{R}}_1$ and
$\Phi^{\text{L}}_2$ denote the right-boundary
projective representation of chain~1 and the left-boundary projective
representation of chain~2, respectively. The label
$\Phi^{\text{I}}_{1\wedge 2}$
denotes the projective stacked representation at the interface
between the two chains.  
}
\end{figure*}
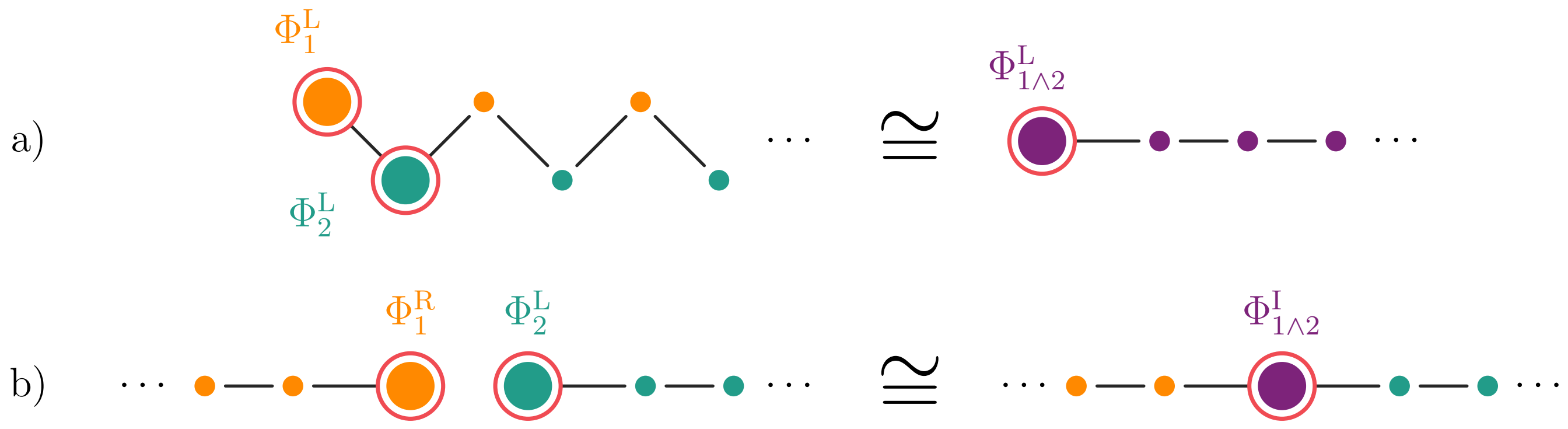

\section{Parafermionic stacking rules}
\label{sec: Parafermionic Stacking Rules}

We now describe the so-called
parafermionic \textit{stacking} operation, that is, a 
composition rule
between two IPT phases that produces a third IPT phase.
We begin by motivating and defining the stacking operation, before computing 
the corresponding \textit{stacking rule} for the classifying indices. 
We do so by distinguishing four cases according 
to the parity of the number of 
boundary parafermions, or in other words, the value of the classifying index $\mu$ for each phase.

We assume $N$ to be a prime power. 
Consider the pair of IPT phases realized by the local Hamiltonians%
~\footnote{
We have assumed that the two Hamiltonians are
defined over chains of the same length $L$.
This assumption is done without loss of generality
after the thermodynamic limit has been taken for each Hamiltonian. }
\begin{subequations}
\label{eq:def for stacking}
\begin{equation}
H^{\,}_{1}:=
\sum^{L}_{j=1}
h^{\,}_{1,j},
\qquad
H^{\,}_{2}:=
\sum^{L}_{j=1}
h^{\,}_{2,j},
\end{equation}
acting on the global Hilbert spaces
$\mathcal{H}^{\,}_{1}$ and $\mathcal{H}^{\,}_{2}$, respectively.
Here, for any $j=1,\cdots,L$ and $a=1,2$,
the energy density $h^{\,}_{a,j}$ is built out of at most $r^{\,}_{a}$
order-$N$ parafermion operators from the subset%
~\footnote{
Here, given $a=1,2$ and $j=1,\cdots,L$, the interval
$I^{\,}_{a,\,j}\subset\{1,\cdots,L\}$
is made of $r^{\,}_{a}$
consecutive sites and contains the site~$j$.  }
\begin{equation}
\big\{
\xi^{\,}_{a,\,k}\equiv\gamma_{a,\,2k-1},
\eta^{\,}_{a,\,k}\equiv\gamma_{a,\,2k}
\big|\
k\in I^{\,}_{a,\,j}
\big\}
\end{equation}  
\end{subequations}
of the associated parafermion algebra 
$\mathfrak{P}^{2L}_{N,a}$.
We associate the triplets of indices 
\begin{align}
\label{eq:left indices stacking}
\Phi^{\mathrm{L}}_{a}
:= 
\bigl(\mu^{\,}_{a},[(\kappa^{\,}_{a},\nu^{\,}_{a})]\bigr)
\end{align}
to the left boundaries of the Hamiltonians $H_a$ 
in Eq.\ \eqref{eq:def for stacking}
when open boundary conditions are imposed, 
where we used $\Phi^{\mathrm{L}}_{a}$ to denote the triplet 
collectively. Recall that consistency demands that the 
triplets of indices associated with the right edge must be
the inverse of the those associated with the left edge,
i.e.,
\begin{align}
\label{eq:right is inverse of left}
\Phi^{\mathrm{R}}_{a}
=
\overline{\Phi^{\mathrm{L}}_{a}},
\end{align}
where the inverse is defined with respect to
the stacking operation we are going to describe.
The IPT phases realized by the Hamiltonians $H_a$
are then fully characterized by the indices 
$\Phi^{\mathrm{L}}_{a}$, $a=1,2$.

To define the stacking operation, we must 
prescribe how the combined parafermionic 
degrees of freedom
\begin{subequations}
\begin{align}
\label{eq:stacked Pf operators}
\big\{
\xi^{\,}_{a,\,k}\equiv\gamma_{a,\,2k-1},
\eta^{\,}_{a,\,k}\equiv\gamma_{a,\,2k}\ \big|\
a=1,2,\
k=1,\cdots,L
\big\}    
\end{align}
act on the stacked Hilbert space
\begin{align}
\mathcal{H}_{1 \wedge 2}
:=
\mathcal{H}_1\otimes\mathcal{H}_2.
\end{align}
This amounts to defining the stacked parafermion algebra 
$\mathfrak{P}^{4L}_{N,1\wedge2}$ and
requires us to choose an ordering
of the parafermion operators \eqref{eq:stacked Pf operators}.
Once this done, we define the stacked Hamiltonian
\begin{equation}
H^{\,}_{1\wedge2}:=
H^{\,}_{1} \otimes \mathbb{1}_{\mathcal{H}_2}
+
\mathbb{1}_{\mathcal{H}_1} \otimes H^{\,}_{2},
\end{equation}
\end{subequations}
acting on the Hilbert space $\mathcal{H}_{1\wedge2}$.

The stacking operation is relevant to two scenarios, shown in Fig.\ 
\ref{fig:two stacking scenarios}, each of which corresponds to a different choice of ordering of the parafermion operators. 

In scenario 1, shown in Fig.\ \ref{fig:two stacking scenarios}(a),
we consider physically putting the two one-dimensional chains that 
support the $\gamma^{\,}_{1,j}$ and $\gamma^{\,}_{2,j}$ degrees of 
freedom, shown in orange and green respectively, on top of each other.
The boundary degrees of freedom realizing the indices 
\eqref{eq:left indices stacking} are schematically marked
with red circles. Crucially, we choose the boundary degrees of freedom
associated to the Hamiltonian $H_1$ to be to the \textit{left} of 
the boundary degrees of freedom associated with Hamiltonian $H_2$.
This choice of ordering will be important. 
An analogous stacking operation 
was first defined in Ref.\ \cite{Motruk2013}.
The stacking results in a new IPT phase,whose degrees of freedom
are shown in purple, is characterized by the left index
\begin{align}
\Phi^{\mathrm{L}}_{1\wedge 2}
=
\Phi^{\text{L}}_1
\wedge
\Phi^{\text{L}}_2.
\end{align}

In scenario 2, shown in Fig.\ \ref{fig:two stacking scenarios}(b),
we consider physically putting the two one-dimensional chains
that support the $\gamma^{\,}_{1,j}$ and $\gamma^{\,}_{2,j}$ degrees of 
freedom, shown in orange and green respectively, next to each other. 
Crucially, we choose all $\gamma^{\,}_{1,j}$ operators
to be to the \textit{left} of all $\gamma^{\,}_{2,j}$
operators. In this case, the stacking does not result in 
a uniform third IPT phase. Instead, it 
results in an interface between the right edge of the IPT phase
described by Hamiltonian $H_1$ and the left edge of the IPT
phase described by Hamiltonian $H_2$. The interface is associated
with the indices
\begin{align}
\Phi^{\text{I}}_{1\wedge2}
=
\Phi^{\text{R}}_1
\wedge
\Phi^{\text{L}}_2
=
\overline{\Phi^{\text{L}}_1}
\wedge
\Phi^{\text{L}}_2,
\end{align}
where we used Eq.\ \eqref{eq:right is inverse of left} in the last
equality. We note that the interface describes a relative index 
between the two IPT phases. Importantly, when
$\Phi^{\text{L}}_1=\Phi^{\text{L}}_2$, 
the index $\Phi^{\text{I}}_{1\wedge2}$ must be trivial
\begin{align}
\Phi^{\text{I}}_{1\wedge2}
=
(0, [(0,0)]),
\end{align}
meaning that the interface can be fully gapped. This is 
expected, since in this case we are just spatially
expanding an IPT phase to the right which should be 
gapped in the bulk. Conversely, when 
$\Phi^{\text{L}}_1\neq\Phi^{\text{L}}_2$,
the interface corresponds to a non-trivial index 
$\Phi^{\text{I}}_{1\wedge2}$. There must
be gapless degrees of freedom localized at the interface
whose degeneracy is described by $\Phi^{\text{I}}_{1\wedge2}$.
If we associate the trivial indices $(0, [(0,0)])$ with the 
vacuum (denoted as \text{vac}), 
then the open boundary conditions can be thought of as an interface between an IPT phase and the vacuum.
In this case, one finds
\begin{align}
\Phi^{\text{I}}_{1\wedge \text{vac}}
=
\Phi^{\text{R}}_1,
\qquad
\Phi^{\text{I}}_{\text{vac} \wedge 2}
=
\Phi^{\text{L}}_2,
\end{align}
which delivers the left and right indices associaed with
and IPT phase, as expected.

In both scenarios described above, we can deduce the staccked indices
$(\mu_{1\wedge2},[(\kappa_{1\wedge 2}, \nu_{1\wedge2})])$
by the indices associated with zero-dimensional parafermionic 
representations of the symmetry group $\mathcal{G}$. 
In particular, the gapped bulk degrees of freedom are immaterial 
to the result of the stacking operation. Without loos of generality,
we assume that the (single-side) boundary degrees of freedom of one stacking component lie either completely to the left or completely to the right of the the (single-side) boundary 
degrees of freedom of the other stacking component.
More precisely, let us denote by the sets for $a=1,2$
\begin{align}
\mathcal{O}^{\,}_{a}
=
\left\{
\gamma^{\,}_{a,1},\,
\gamma^{\,}_{a,2},
\cdots
\gamma^{\,}_{a,n_a},
\right\}
\end{align}
The boundary degrees of freedom that realize the projective 
representations $U^{\,}_a(\tilde{g})$ of $\mathcal{G}$.
We denote by $1\wedge2$ the stacking operation where the entirety
of the boundary degrees of freedom $\mathcal{O}^{\,}_{1}$ are
chosen to be the \textit{left} 
of the entirety of the boundary degrees
of freedom $\mathcal{O}^{\,}_{2}$ (the opposite case being $2\wedge 1$).

We then define the stacked representation 
$U^{\,}_{1\wedge2}$ of the symmetry group $\mathcal{G}$ 
by demanding that it acts on each stacking-component $\mathcal{O}^{\,}_{a}$
in accordance with its respective representation $U^{\,}_a$, 
as expressed by the following consistency equations 
\begin{subequations}
\label{eq:stacked rep key relation}
\begin{align}
&
U^{\,}_{1\wedge2}(\tilde{g})\,
\gamma^{\,}_{1,\,j}\,
U^{\dag}_{1\wedge2}(\tilde{g})=
U^{\,}_{1}(\tilde{g})\,
\gamma^{\,}_{1,\,j}\,
U^{\dag}_{1}(\tilde{g}),
\\
&
U^{\,}_{1\wedge2}(\tilde{g})\,
\gamma^{\,}_{2,\,k}\,
U^{\dag}_{1\wedge2} (\tilde{g})=
U^{\,}_{2}(\tilde{g})\,
\gamma^{\,}_{2,\,k}\,
U^{\dag}_{2}(\tilde{g}),
\end{align}
\end{subequations}
where $j=1,\cdots,n_1$ and $k=1,\cdots, n_2$.

In what follows, we consider the stacking for 
four cases where $\mu_1=0,1$ and $\mu_2=0,1$. 
In each case, we denote by $Q_{1\wedge 2}$,
$C_{1\wedge 2}(g)$, $Y_{1\wedge 2}$ the 
stacked parafermion charge operator, the charge conjugation
operator associate with element $g\in G_b$,
and the generator of the center when the stacked index
$\mu_{1\wedge2}=1$, according to the definitions introduced
in Sec.\ \ref{sec:Boundary projective representations}.

\subsection{Even-even stacking}
\label{subsec:ee stacking}

For even-even stacking, we have
$\mu^{\,}_{1}=\mu^{\,}_{2}=0$. 
We thus compute the stacked index
\begin{align}
\mu^{\,}_{1\wedge2}=
\mu^{\,}_{1}
+
\mu^{\,}_{2}=
0 \mod 2.
\end{align}
The boundary projective representations
$U^{\,}_{1}(z,g)$ and $U^{\,}_{2}(z,g)$ 
are of the form \eqref{eq:general proj rep mu even}.
We define for any $(z,g)\in\mathcal{G}$ the stacked representation ansatz
\begin{subequations}
\label{eq:stacked rep ee}
\begin{align}
U^{\,}_{1\wedge2}(z,g):=
Q^{z}_{1\wedge2}\,
C^{\,}_{{1\wedge2}}(g)\,
V^{\,}_{{1\wedge2}}(g),
\label{eq:stacked rep ee a}
\end{align}
where 
\begin{align}
V^{\,}_{{1\wedge2}}(g):=
V^{\,}_{1}(g)\,
V^{\,}_{2}(g)\,
Q^{-\kappa^{\,}_{2}(g)}_{1}\,
Q^{+\kappa^{\,}_{1}(g)}_{2}.
\label{eq:stacked rep ee b}
\end{align}
\end{subequations}
One verifies that the stacked representation \eqref{eq:stacked rep ee}
satisfies the consistency relations \eqref{eq:stacked rep key relation}. 
In particular, the parafermion charge operators
$Q^{\,}_{1}$ and $Q^{\,}_{2}$ 
are needed in Eq.\ \eqref{eq:stacked rep ee b} to compensate for 
the multiplicative phase
$\omega^{-\kappa^{\,}_{1}(g)}_{N}$
that appears when conjugating $\gamma^{\,}_{2,k}$ by
$V^{\,}_{1}(g)$ 
and the multiplicative phase
$\omega^{+\kappa^{\,}_{2}(g)}_{N}$ that appears
when conjugating $\gamma^{\,}_{1,j}$ by $V^{\,}_{2}(g)$. 

To compute the indices $\kappa^{\,}_{1\wedge2}$ and $\nu^{\,}_{1\wedge2}$ 
associated with the stacked representation \eqref{eq:stacked rep ee},
it is sufficient to consider the composition of 
two elements $\tilde{g}\equiv(z,g)$ and 
$\tilde{h}\equiv(u,h)$ in $\mathcal{G}$, i.e.,
\begin{align}
U^{\,}_{1\wedge2}(\tilde{g})\,
U^{\,}_{1\wedge2}(\tilde{h})
&=
Q^{z}_{1\wedge2}\,
C^{\,}_{{1\wedge2}}(g)\,
V^{\,}_{{1\wedge2}}(g)
\nonumber\\
&\times
Q^{u}_{1\wedge2}\,
C^{\,}_{{1\wedge2}}(h)\,
V^{\,}_{{1\wedge2}}(h).
\end{align}
\normalsize
Rearranging the terms produces phase factors dependent on the parafermion
charges of each operator according to Eq.\
\eqref{eq:algebra well separated Pf ops}.
The operators $Q^{\,}_{1}$ and
$Q^{\,}_{2}$ carry no parafermion charge, while the
operators $V^{\,}_{1}(g)$ and $V^{\,}_{2}(g)$
carry parafermion charges $-\kappa^{\,}_{1}(g)$ and $-\kappa^{\,}_{2}(g)$,
respectively. The charge conjugation operator $C^{\,}_{{1\wedge2}}(g)$
does not have a definite parafermion charge.
Finally, the parafermion charge of the operator
\begin{align}
\label{eq:charge conj shorthand}
\left[V^{\,}_{a}(g)\right]^{\rho_{h}}\:=
C^{\,}_{{1\wedge2}}(h)\,
V^{\,}_{a}(g)\,
C^{\,}_{{1\wedge2}}(h),
\end{align}
is simply $-\rho_{h}\,\kappa^{\,}_{a}(g)$, with  $a=1,2$.
After rearranging the terms, we obtain
\begin{align}
U^{\,}_{1\wedge2}(\tilde{g})\,
U^{\,}_{1\wedge2}(\tilde{h})&=
\omega^{
[
\kappa^{\,}_{1}(g)
+
\kappa^{\,}_{2}(g)
]\,u
-
\rho_{h}\,
\kappa^{\,}_{1}(g)\,
\kappa^{\,}_{2}(h)
}_{N}
\nonumber\\
&\times
Q^{z+\rho_{g}u}_{1\wedge2}\,
C^{\,}_{{1\wedge2}}(g\,h)
\nonumber\\
&\times
\left[V^{\,}_{1}(g)\right]^{\rho_{h}}\,V^{\,}_{1}(h)\,
\left[V^{\,}_{2}(g)\right]^{\rho_{h}}\,V^{\,}_{2}(h)
\nonumber\\
&\times
Q^{-\rho_{h}\kappa^{\,}_{2}(g)-\kappa^{\,}_{2}(h)}_{1}\,
Q^{+\rho_{h}\kappa^{\,}_{1}(g)+\kappa^{\,}_{1}(h)}_{2}.
\end{align}

Now, we  use the following two identities for $a=1,2$.
First,
\begin{subequations}
\begin{align}
\label{eq:composition identity even}
\left[V^{\,}_{a}(g)\right]^{\rho_{h}}\,V^{\,}_{a}(h)=
e^{\mathrm{i}\nu^{\,}_{a}(g,h)}\,
Q^{\rho_{gh}c(g,h)}_{a}\,
V^{\,}_{a}(g\,h),
\end{align}
which follows immediately from Eq.\ \eqref{eq:index nu mu even}. 
Second,
\begin{align}
\label{eq:cocylce cond on kappa}
\rho_{h}\kappa^{\,}_{a}(g)+\kappa^{\,}_{a}(h)= \kappa^{\,}_{a}(g\,h),
\end{align}
\end{subequations}
which follows from the cocycle condition $\delta^{(1)}_{\rho}\kappa = 0$.
These two identities deliver
\begin{align}
U^{\,}_{1\wedge2}(\tilde{g})\,
U^{\,}_{1\wedge2}(\tilde{h})&=
\omega^{[\kappa^{\,}_{1}(g)+\kappa^{\,}_{2}(g)]u}_{N}
\nonumber\\
&\times
e^{
\mathrm{i}\nu^{\,}_{1}(g,h)
+
\mathrm{i}\nu^{\,}_{2}(g,h)
-
\mathrm{i}\frac{2\pi}{N}\,
\rho_{h}\,\kappa^{\,}_{1}(g)\,\kappa^{\,}_{2}(h)
}
\nonumber \\
&\times
Q^{z+\rho_{g}u+c(g,h)}_{1\wedge2}\,
C^{\,}_{{1\wedge2}}(g\,h)
\nonumber\\
&\times
V^{\,}_{1}(g\,h)\,
V^{\,}_{2}(g\,h)\,
Q^{-\kappa^{\,}_{2}(g\,h)}_{1}\,
Q^{+\kappa^{\,}_{1}(g\,h)}_{2}
\nonumber\\
&=
\omega^{[\kappa^{\,}_{1}(g)+\kappa^{\,}_{2}(g)]u}_{N}
\nonumber\\
&\times
e^{
\mathrm{i}\nu^{\,}_{1}(g,h)
+
\mathrm{i}\nu^{\,}_{2}(g,h)
-
\mathrm{i}\frac{2\pi}{N}\,
\rho_{h}\,
\kappa^{\,}_{1}(g)\,
\kappa^{\,}_{2}(h)
}
\nonumber\\
&\times
U^{\,}_{1\wedge2}(\tilde{g}\circ \tilde{h}).
\end{align}
Comparison with Eq.\ \eqref{eq:phi in terms of nu kappa mu even} 
yields the stacked indices
\begin{subequations}
\label{eq:stacked indices ee}
\begin{align}
&
\kappa^{\,}_{1\wedge2}(g)=
\kappa^{\,}_{1}(g)
+
\kappa^{\,}_{2}(g),
\\
&
\nu^{\,}_{1\wedge2}(g,h)=
\nu^{\,}_{1}(g,h)+\nu^{\,}_{2}(g,h)
-
\frac{2\pi}{N}\,
\rho_{h}\,
\kappa^{\,}_{1}(g)\,
\kappa^{\,}_{2}(h),
\end{align}
\end{subequations}
which we write compactly as
\begin{subequations}
\label{eq:stacked indices ee bis}
\begin{align}
&
\kappa^{\,}_{1\wedge2} =
\kappa^{\,}_{1}
+
\kappa^{\,}_{2},
\\
&
\nu^{\,}_{1\wedge2}=
\nu^{\,}_{1}
+
\nu^{\,}_{2}
-
\frac{2\pi}{N}\,
\kappa^{\,}_{1}\cup\kappa^{\,}_{2}.
\end{align}
\end{subequations}
The index $\kappa^{\,}_{1\wedge 2}$ quantifies the combined
parafermion charge of the unitary operators $V^{\,}_{1}$ and
$V^{\,}_{2}$.
The stacked index $\nu^{\,}_{1\wedge2}$ acquires an additional contribution
$\frac{2\pi}{N}\kappa^{\,}_{1}\cup\kappa^{\,}_{2}$ due to the
parafermionic nature of the underlying degrees of freedom. When $N=2$,
this additional contribution reduces to the case of Majorana degrees
of freedom~\cite{Aksoy2022}. In contrast to the Majorana case, the
stacking rule \eqref{eq:stacked indices ee} can lead to a
\textit{non-Abelian} stacking group of IPT phases, see Sec.\ \ref{sec:stacking example} for such an example.
This is because for any two 1-cocycles
$\kappa,\,\kappa'\in Z^{1}(G^{\,}_{b},[\Z^{\mathrm{pf}}_{N}]^{\,}_{\rho})$, one has the identity
\begin{subequations}
\begin{align}
\label{eq:useful cocycle identity}
\rho_{h}
\kappa(g)\,
\kappa'(h)
\sim
-
\rho_{h}\,
\kappa'(g)\,
\kappa(h).
\end{align}
In other words, the cup product of the two $1$-cocycles
is antisymmetric up to a $2$-coboundary, i.e.,
\begin{equation}
\kappa^{\,}_{1}\cup\kappa^{\,}_{2} 
\sim
-\kappa^{\,}_{2}\cup\kappa^{\,}_{1}
\Longrightarrow
\nu^{\,}_{1\wedge2}\neq
\nu^{\,}_{2\wedge1}.
\label{eq:non-commutative cup product}
\end{equation}   
\end{subequations}

\subsection{Odd-odd stacking}
\label{subsec:oo stacking}

For odd-odd stacking, we have both $\mu^{\,}_{1}=\mu^{\,}_{2}=1$. 
We thus compute the stacked index
\begin{align}
\mu^{\,}_{1\wedge2}=
\mu^{\,}_{1}
+
\mu^{\,}_{2}=
0 \mod 2.
\end{align}
The boundary projective representations
$U^{\,}_{1}(z,g)$ and $U^{\,}_{2}(z,g)$ 
are of the form \eqref{eq:general proj rep mu odd}. 
Since $\mu^{\,}_{1\wedge2}=0$, the stacked representation must act on a
local Hilbert space and must be built out of $n^{\,}_{1}+n^{\,}_{2}$
local parafermion operators.
In other words, the stacked representation cannot contain
the auxiliary parafermions
$\gamma_{1,\infty}$ and $\gamma_{2,\infty}$. 
As we shall see, the key idea is that the central elements
$Y^{\,}_{2}$ and $Y^{\,}_{1}$
act just like  
$\gamma^{\,}_{1,\infty}$ and $\gamma^{\dag}_{2,\infty}$
on the parafermion operators 
$\gamma^{\,}_{1,j}$ and $\gamma^{\,}_{2,k}$, respectively.

We define for any $(z,g)\in \mathcal{G}$ the stacked representation ansatz
\begin{subequations}
\label{eq:stacked rep oo}
\begin{align}
\label{eq:stacked rep oo a}
U^{\,}_{1\wedge2}(z,g):=
Q^{z}_{1\wedge2}\,
C^{\,}_{{1\wedge2}}(g)\,
V^{\,}_{{1\wedge2}}(g),
\end{align}
where 
\begin{align}
\label{eq:stacked rep oo b}
V^{\,}_{{1\wedge2}}(g):=
V^{\,}_{1}(g)\,
V^{\,}_{2}(g)\,
Y^{+\kappa^{\,}_{2}(g)}_{1}\,
Y^{-\kappa^{\,}_{1}(g)}_{2}.
\end{align}
\end{subequations}
One verifies that the stacked representation \eqref{eq:stacked rep oo}
satisfies the consistency relations \eqref{eq:stacked rep key relation}. 
Note that, by definition, the operators
$V^{\,}_{1}(g)$ and $V^{\,}_{2}(g)$
do not carry parafermion charges. 
Hence, for any $g$,
the product
$V^{\,}_{1}(g)\,Y^{+\kappa^{\,}_{1}(g)}_{2}$
commutes with all $\gamma^{\,}_{2,k}$ parafermions.
Similarly, the product
$V^{\,}_{2}(g)\,Y^{-\kappa^{\,}_{2}(g)}_{1}$
commutes with all
$\gamma^{\,}_{1,j}$ parafermions.

To compute the indices $\kappa^{\,}_{1\wedge2}$ and $\nu^{\,}_{1\wedge2}$ 
associated with the representation \eqref{eq:stacked rep oo},
it is sufficient to consider the composition of 
two elements $\tilde{g}\equiv(z,g)$ and 
$\tilde{h}\equiv(u,h)$ in $\mathcal{G}$, i.e.,
\begin{align}
U^{\,}_{1\wedge2}(\tilde{g})\,
U^{\,}_{1\wedge2}(\tilde{h})
&=
Q^{z}_{1\wedge2}\,
C^{\,}_{{1\wedge2}}(g)\,
V^{\,}_{{1\wedge2}}(g)
\nonumber\\
&\times
Q^{u}_{1\wedge2}\,
C^{\,}_{{1\wedge2}}(h)\,
V^{\,}_{{1\wedge2}}(h).
\end{align}
Upon reordering of the terms, we find
\begin{align}
U^{\,}_{1\wedge2}(\tilde{g})\,
U^{\,}_{1\wedge2}(\tilde{h})
&=
\omega^{
[
\kappa^{\,}_{1}(g)
-
\kappa^{\,}_{2}(g)
]\,u
+
\rho_{h}\,
\kappa^{\,}_{1}(g)\,
\kappa^{\,}_{2}(h)
}_{N}
\nonumber\\
&\times
Q^{z+\rho_{g}\,u}_{1\wedge2}\,
C^{\,}_{{1\wedge2}}(g\,h)
\nonumber\\
&\times
[V^{\,}_{1}(g)]^{\rho_{h}}
V^{\,}_{1}(h)\,
[V^{\,}_{2}(g)]^{\rho_{h}}
V^{\,}_{2}(h)
\nonumber\\
&
\times
Y^{+\rho_{h}\,\kappa^{\,}_{2}(g)+\kappa^{\,}_{2}(h)}_{1}\,
Y^{-\rho_{h}\,\kappa^{\,}_{1}(g)-\kappa^{\,}_{1}(h)}_{2},
\end{align}
where the multiplicative phase factors arise
when we pass $Q^{u}_{1\wedge2}$ through
$Y^{\,}_{1}$ 
and
$Y^{\,}_{2}$, and when
$Y^{-\rho_{h}\kappa^{\,}_{1}(g)}_{2}$
is passed through
$Y^{+\kappa^{\,}_{2}(h)}_{1}$.

Using the identity following from Eq.\ \eqref{eq:index nu mu odd}
\begin{align}
\label{eq:composition identity odd}
[V^{\,}_{a}(g)]^{\rho_{h}}
V^{\,}_{a}(h)=
e^{\mathrm{i}\nu^{\,}_{a}(g,h)}\,
V^{\,}_{a}(g\,h),
\end{align}
in combination with the cocycle condition \eqref{eq:cocylce cond on kappa} delivers the final result,
\begin{align}
U^{\,}_{1\wedge2}(\tilde{g})\,
U^{\,}_{1\wedge2}(\tilde{h})
&
=
\omega^{[\kappa^{\,}_{1}(g)-\kappa^{\,}_{2}(g)]u}_{N}
\nonumber\\
&\times
e^{\mathrm{i}\nu^{\,}_{1}(g,h) + \mathrm{i}\nu^{\,}_{2}(g,h)
+ \mathrm{i}\frac{2\pi}{N}\rho_{h}\kappa^{\,}_{1}(g)\,\kappa^{\,}_{2}(h)}
\nonumber\\
&\times
Q^{z+\rho_{g}u}_{1\wedge2}\,
C^{\,}_{{1\wedge2}}(g\,h)
\nonumber\\
&\times
V^{\,}_{1}(g\,h)\,
V^{\,}_{2}(g\,h)\,
Y^{+\kappa^{\,}_{2}(g\,h)}_{1}\,
Y^{-\kappa^{\,}_{1}(g\,h)}_{2}
\nonumber\\
&=
\omega^{[\kappa^{\,}_{1}(g)-\kappa^{\,}_{2}(g)]u}_{N}
\nonumber\\
&\times
e^{\mathrm{i}\nu^{\,}_{1}(g,h) + \mathrm{i}\nu^{\,}_{2}(g,h)
+ \mathrm{i}\frac{2\pi}{N}\rho_{h}\kappa^{\,}_{1}(g)\,\kappa^{\,}_{2}(h)}
\nonumber\\
&\times
U^{\,}_{1\wedge2}(\tilde{g}\circ \tilde{h}).
\end{align}
Comparison with Eq.\ \eqref{eq:phi in terms of nu kappa mu even} 
yields the stacked indices
\begin{subequations}
\label{eq:stacked indices oo}
\begin{align}
&
\kappa^{\,}_{1\wedge2}(g)=
\kappa^{\,}_{1}(g)
-
\kappa^{\,}_{2}(g),
\\
&
\nu^{\,}_{1\wedge2}(g,h)=
\nu^{\,}_{1}(g,h)+\nu^{\,}_{2}(g,h) +
\frac{2\pi}{N}\,
\rho_{h}\,
\kappa^{\,}_{1}(g)\,
\kappa^{\,}_{2}(h),
\end{align}
\end{subequations}
which we write compactly as
\begin{subequations}
\label{eq:stacked indices oo bis}
\begin{align}
&
\kappa^{\,}_{1\wedge2} =
\kappa^{\,}_{1}
-
\kappa^{\,}_{2},
\\
&
\nu^{\,}_{1\wedge2}=
\nu^{\,}_{1}
+
\nu^{\,}_{2}
+
\frac{2\pi}{N}\,\kappa^{\,}_{1}\cup\kappa^{\,}_{2}.
\end{align}
\end{subequations}
The index $\kappa^{\,}_{1\wedge 2}$
quantifies the total parafermion charge
of the unitary operator $V^{\,}_{1\wedge2}$. 
The stacked index $\nu^{\,}_{1\wedge2}$
acquires an additional contribution 
$\frac{2\pi}{N}\kappa^{\,}_{1}\cup\kappa^{\,}_{2}$
due to the parafermionic nature of 
the underlying degrees of freedom.
Note the sign differences compared to 
the even-even stacking case \eqref{eq:stacked indices ee}.
Here too, the non-commutativity
(\ref{eq:non-commutative cup product})
holds.

\subsection{Even-odd stacking}
\label{subsec:eo stacking}

For even-odd stacking, we have $\mu^{\,}_{1}=0$ and $\mu^{\,}_{2}=1$. 
We thus compute the stacked index
\begin{align}
\mu^{\,}_{1\wedge2} = \mu^{\,}_{1} + \mu^{\,}_{2} = 1 \mod 2.
\end{align}
The boundary projective representations
$U^{\,}_{1}(z,g)$ and $U^{\,}_{2}(z,g)$ 
are of the forms  \eqref{eq:general proj rep mu even}
and \eqref{eq:general proj rep mu odd}, respectively. 
The stacked representation must be of the form
\eqref{eq:general proj rep mu odd},
where the auxiliary parafermion is supplied by the operator
$\gamma^{\,}_{2,\infty}$.
Hence, we identify
$\gamma^{\,}_{1\wedge 2,\infty}\equiv\gamma^{\,}_{2,\infty}$.

We define for any $(z,g)\in \mathcal{G}$ the stacked 
representation ansatz
\begin{subequations}
\label{eq:stacked rep eo}
\begin{align}
\label{eq:stacked rep eo a}
U^{\,}_{1\wedge2}(z,g):=
Q^{z}_{1\wedge2}\,
C^{\,}_{{1\wedge2}}(g)\,
V^{\,}_{{1\wedge2}}(g)\,
\gamma^{-(\kappa^{\,}_{1}(g)+\kappa^{\,}_{2}(g))}_{1\wedge 2, \infty},
\end{align}
where 
\begin{align}
\label{eq:stacked rep eo b}
V^{\,}_{{1\wedge2}}(g):=
V^{\,}_{1}(g)\,
V^{\,}_{2}(g)\,
Q^{-\kappa^{\,}_{2}(g)}_{1}\,
Y^{+\kappa^{\,}_{1}(g)}_{2}.
\end{align}
\end{subequations}
One verifies that the stacked representation
\eqref{eq:stacked rep oo}
satisfies the consistency relations
\eqref{eq:stacked rep key relation}. 
Note that the operator $V^{\,}_{{1\wedge2}}(g)$ carries
no parafermion charge. This is because both $V^{\,}_{2}(g)$
and $Q^{\,}_{1}$
carry no parafermion charges,
while the parafermion charge
of $Y^{+\kappa^{\,}_{1}(g)}_{2}$ compensates that of the operator
$V^{\,}_{1}(g)$.
Similarly, one observes that the operator $V^{\,}_{{1\wedge2}}(g)$
commutes with both the central element 
$Y^{\,}_{1\wedge 2}$ and 
the auxiliary parafermion
$\gamma^{\,}_{1\wedge 2, \infty}$.

Comparing with Eq.\ \eqref{eq:general proj rep mu odd},
one finds that the stacked index
$\kappa^{\,}_{1\wedge2}$
is  given by
\begin{align}
\kappa^{\,}_{1\wedge2}(g)=
\kappa^{\,}_{1}(g)
+
\kappa^{\,}_{2}(g).
\end{align}
The stacked index $\nu^{\,}_{1\wedge2}$ can then be obtained by computing
\begin{align}
U^{\,}_{1\wedge2}(0,g)\,
U^{\,}_{1\wedge2}(0,h)&=
C^{\,}_{{1\wedge2}}(g)\,
V^{\,}_{{1\wedge2}}(g)\,
\gamma^{-\kappa^{\,}_{1\wedge2}(g)}_{1\wedge 2,\infty}
\nonumber\\
&\times
C^{\,}_{{1\wedge2}}(h)\,
V^{\,}_{{1\wedge2}}(h)\,
\gamma^{-\kappa^{\,}_{1\wedge2}(h)}_{1\wedge 2,\infty}
\nonumber\\
&= 
C^{\,}_{{1\wedge2}}(g\,h)\,
[V^{\,}_{{1\wedge2}}(g)]^{\rho_{h}}\,
V^{\,}_{{1\wedge2}}(h)
\nonumber\\
&\times
\gamma^{-\rho_{h}\kappa^{\,}_{1\wedge2}(g)-\kappa^{\,}_{1\wedge2}(h)}_{1\wedge 2,\infty},
\end{align}
where in reaching the last line we used
the fact that $V^{\,}_{{1\wedge2}}(g)$
commutes with 
$\gamma^{\,}_{1\wedge 2, \infty}$.
We now rearrange the terms in the product
$[V^{\,}_{{1\wedge2}}(g)]^{\rho_{h}}$
as
\small
\begin{align}
[V^{\,}_{{1\wedge2}}(g)]^{\rho_{h}}\,
V^{\,}_{{1\wedge2}}(h)&=
[V^{\,}_{1}(g)]^{\rho_{h}}\,
[V^{\,}_{2}(g)]^{\rho_{h}}\,
Q^{-\rho_{h}\kappa^{\,}_{2}(g)}_{1}\,
Y^{+\rho_{h}\kappa^{\,}_{1}(g)}_{2}
\nonumber\\
&\times
V^{\,}_{1}(h)\,
V^{\,}_{2}(h)\,
Q^{-\kappa^{\,}_{2}(h)}_{1}\,
Y^{+\kappa^{\,}_{1}(h)}_{2}
\nonumber\\
&=
\omega^{
\rho_{h}\,
\kappa^{\,}_{1}(g)\,
\kappa^{\,}_{1}(h)
+
\rho_{h}\,
\kappa^{\,}_{2}(g)\,
\kappa^{\,}_{1}(h)
}_{N}
\nonumber\\
&\times
e^{\mathrm{i}\nu^{\,}_{1}(g,h)+\mathrm{i}\nu^{\,}_{2}(g,h)}\,
\nonumber\\
&\times
V^{\,}_{1}(g\,h)\,
V^{\,}_{2}(g\,h)\,
Q^{-\kappa^{\,}_{2}(g\,h)}_{1}\,
Y^{+\kappa^{\,}_{1}(g\,h)}_{2},
\end{align}
\normalsize
where we used Eq.\ \eqref{eq:composition identity even} with $c=0$ for 
the unitary $V^{\,}_{1}$
and Eq.\
\eqref{eq:composition identity odd}
for  the unitary $V^{\,}_{2}$.%
~\footnote{
Recall that, since we are stacking an odd representation,
the extension class must be trivial, i.e., $c=0$.}
The last line is precisely the unitary
$V^{\,}_{1\wedge 2}(g\,h)$.
We thus find 
\begin{subequations}
\begin{align}
U^{\,}_{1\wedge2}(0,g)\,
U^{\,}_{1\wedge2}(0,h)&= 
\omega^{
\rho_{h}\,
\kappa^{\,}_{1}(g)\,
\kappa^{\,}_{1}(h)
+
\rho_{h}\,
\kappa^{\,}_{2}(g)\,
\kappa^{\,}_{1}(h)
}_{N}
\nonumber\\
&\times
e^{\mathrm{i}\nu^{\,}_{1}(g,h)+\mathrm{i}\nu^{\,}_{2}(g,h)}\,
\nonumber\\
&\times
U^{\,}_{1\wedge2}(0,g\,h),
\end{align}
and the stacked index $\nu^{\,}_{1\wedge 2}(g,h)$
is given by
\begin{align}
\nu^{\,}_{1\wedge 2}(g,h)&=
\nu^{\,}_{1}(g,h)
+
\nu^{\,}_{2}(g,h)
\nonumber\\
&
+
\frac{2\pi}{N}
\rho_{h}
\left(
\kappa^{\,}_{1}(g)\,\kappa^{\,}_{1}(h)
+
\kappa^{\,}_{2}(g)\,\kappa^{\,}_{1}(h)
\right).
\end{align}
\end{subequations}
Here, we can simplify the contributions from the 1-cocycles
$\kappa^{\,}_{1}$
and
$\kappa^{\,}_{2}$ by using the identity \eqref{eq:useful cocycle identity}
to finally obtain
\begin{align}
\nu^{\,}_{1\wedge 2}(g,h)=
\nu^{\,}_{1}(g,h)
+
\nu^{\,}_{2}(g,h)
-
\frac{2\pi}{N}\,
\rho_{h}\,
\kappa^{\,}_{1}(g)\,
\kappa^{\,}_{2}(h).
\end{align}
We write the stacked indices compactly as
\begin{subequations}
\label{eq:stacked indices eo}
\begin{align}
&
\kappa^{\,}_{1\wedge2}=
\kappa^{\,}_{1}
+
\kappa^{\,}_{2},
\\
&
\nu^{\,}_{1\wedge2}=
\nu^{\,}_{1}
+
\nu^{\,}_{2}
-
\frac{2\pi}{N}\,
\kappa^{\,}_{1}\cup\kappa^{\,}_{2}.
\end{align}
\end{subequations}
Here too, the non-commutativity
(\ref{eq:non-commutative cup product})
holds.

\subsection{Odd-even stacking}
\label{subsec:oe stacking}

For odd-even stacking, we have $\mu^{\,}_{1}=1$ and $\mu^{\,}_{2}=0$. 
We thus compute the stacked index
\begin{align}
\mu^{\,}_{1\wedge2}=
\mu^{\,}_{1}
+
\mu^{\,}_{2}=
1 \mod 2.
\end{align}
The boundary projective representations
$U^{\,}_{1}(z,g)$ and $U^{\,}_{2}(z,g)$ 
are of the forms \eqref{eq:general proj rep mu odd}
and \eqref{eq:general proj rep mu even}, respectively. 
The stacked representation must be of the form
\eqref{eq:general proj rep mu odd}
where the auxiliary parafermion is supplied by the operator
$\gamma^{\,}_{1,\infty}$.
Hence, we identify
$\gamma^{\,}_{1\wedge 2,\infty}\equiv
\gamma^{\,}_{1,\infty}$.
Recall that in the even-odd stacking case of Sec.\
\ref{subsec:eo stacking}, 
the operators $\gamma^{\,}_{2,k}$
realizing an odd parafermion algebra lay to the ``right''
of those realizing an even parafermion algebra. 
In contrast, in odd-even stacking the operators $\gamma^{\,}_{1,j}$  
realizing an odd parafermion algebra lie to the ``left''
of those realizing an even parafermion algebra.
This will lead to different  stacking rules.

We define for any $(z,g)\in \mathcal{G}$ the stacked representation ansatz
\begin{subequations}
\label{eq:stacked rep oe}
\begin{align}
\label{eq:stacked rep oe a}
U^{\,}_{1\wedge2}(z,g):=
Q^{z}_{1\wedge2}\,
C^{\,}_{{1\wedge2}}(g)\,
V^{\,}_{{1\wedge2}}(g)\,
\gamma^{-(\kappa^{\,}_{1}(g)-\kappa^{\,}_{2}(g))}_{1\wedge 2,\infty},
\end{align}
where 
\begin{align}
\label{eq:stacked rep oe b}
V^{\,}_{{1\wedge2}}(g):=
V^{\,}_{1}(g)\,
V^{\,}_{2}(g)\,
Q^{-\kappa^{\,}_{1}(g)}_{2}\,
Y^{+\kappa^{\,}_{2}(g)}_{1}.
\end{align}
\end{subequations}
One verifies that the stacked representation \eqref{eq:stacked rep oo}
satisfies the consistency relations \eqref{eq:stacked rep key relation}. 
Note that the operator $V^{\,}_{{1\wedge2}}(g)$ carries
no parafermion charge. This is because both
$V^{\,}_{1}(g)$
and
$Q^{\,}_{1}$
carry no parafermion charges, while the parafermion charge of
$Y^{+\kappa^{\,}_{2}(g)}_{1}$
compensates that of the operator
$V^{\,}_{2}(g)$.
Similarly, one observes that the operator
$V^{\,}_{{1\wedge2}}(g)$
commutes
with both the central element 
$Y^{\,}_{1\wedge 2}$ and 
the auxiliary parafermion $\gamma^{\,}_{1\wedge 2, \infty}$.

Comparing with Eq.\ \eqref{eq:general proj rep mu odd},
one finds that the 
stacked index $\kappa^{\,}_{1\wedge2}$ is given by
\begin{align}
\kappa^{\,}_{1\wedge2}(g)=
\kappa^{\,}_{1}(g)
-
\kappa^{\,}_{2}(g).
\end{align}
The stacked index $\nu^{\,}_{1\wedge2}$ can then be obtained by computing
\small
\begin{align}
U^{\,}_{1\wedge2}(0,g)\,
U^{\,}_{1\wedge2}(0,h)
&=
C^{\,}_{{1\wedge2}}(g)\,
V^{\,}_{{1\wedge2}}(g)\,
\gamma^{-\kappa^{\,}_{1\wedge2}(g)}_{1\wedge 2,\infty}
\nonumber\\
&
\times
C^{\,}_{{1\wedge2}}(h)\,
V^{\,}_{{1\wedge2}}(h)\,
\gamma^{-\kappa^{\,}_{1\wedge2}(h)}_{1\wedge 2,\infty}
\nonumber\\
&= 
C^{\,}_{{1\wedge2}}(g\,h)\,
[V^{\,}_{{1\wedge2}}(g)]^{\rho_{h}}\,
V^{\,}_{{1\wedge2}}(h)
\nonumber\\
&\times
\gamma^{-\rho_{h}\kappa^{\,}_{1\wedge2}(g)-\kappa^{\,}_{1\wedge2}(h)}_{1\wedge 2,\infty},
\end{align}
\normalsize
where in reaching the last line we used
the fact that $V^{\,}_{{1\wedge2}}(g)$ commutes with 
$\gamma^{\,}_{1\wedge 2, \infty}$.
We now rearrange the terms in the product
$[V^{\,}_{{1\wedge2}}(g)]^{\rho_{h}}$
as
\small
\begin{align}
[V^{\,}_{{1\wedge2}}(g)]^{\rho_{h}}\,
V^{\,}_{{1\wedge2}}(h)&=
[V^{\,}_{1}(g)]^{\rho_{h}}\,
[V^{\,}_{2}(g)]^{\rho_{h}}\,
Q^{-\rho_{h}\kappa^{\,}_{1}(g)}_{2}\,
Y^{+\rho_{h}\kappa^{\,}_{2}(g)}_{1}
\nonumber\\
&\times
V^{\,}_{1}(h)\,
V^{\,}_{2}(h)\,
Q^{-\kappa^{\,}_{1}(h)}_{2}\,
Y^{+\kappa^{\,}_{2}(h)}_{1}
\nonumber\\
&=
\omega^{
\rho_{h}\,
\kappa^{\,}_{2}(g)\,
\kappa^{\,}_{2}(h)
+
\rho_{h}\,
\kappa^{\,}_{1}(g)\,
\kappa^{\,}_{2}(h)
}_{N}
\nonumber\\
&\times
e^{\mathrm{i}\nu^{\,}_{1}(g,h)+\mathrm{i}\nu^{\,}_{2}(g,h)}\,
\nonumber\\
&\times
V^{\,}_{1}(g\,h)\,
V^{\,}_{2}(g\,h)\,
Q^{-\kappa^{\,}_{1}(g\,h)}_{2}\,
Y^{+\kappa^{\,}_{2}(g\,h)}_{1},
\end{align}
\normalsize
where we used Eq.\ \eqref{eq:composition identity even}
with $c=0$ for  the unitary $V^{\,}_{1}$ and Eq.\
\eqref{eq:composition identity odd}
for  the unitary $V^{\,}_{2}$.%
~\footnote{
Recall that, since we are stacking an odd representation,
the extension class must be trivial, i.e., $c=0$.}
The last line is precisely the unitary $V^{\,}_{1\wedge 2}(g\,h)$.
We thus find 
\begin{subequations}
\begin{align}
U^{\,}_{1\wedge2}(0,g)\,
U^{\,}_{1\wedge2}(0,h)&= 
\omega^{
\rho_{h}\,
\kappa^{\,}_{2}(g)\,
\kappa^{\,}_{2}(h)
-
\rho_{h}\,
\kappa^{\,}_{1}(g)\,
\kappa^{\,}_{2}(h)
}_{N}
\nonumber\\
&\times
e^{\mathrm{i}\nu^{\,}_{1}(g,h)+\mathrm{i}\nu^{\,}_{2}(g,h)}\,
\nonumber\\
&\times
U^{\,}_{1\wedge2}(0,g\,h),
\end{align}
and the stacked index $\nu^{\,}_{1\wedge 2}(g,h)$
is given by
\begin{align}
\nu^{\,}_{1\wedge 2}(g,h)&=
\nu^{\,}_{1}(g,h)
+
\nu^{\,}_{2}(g,h)
\nonumber\\
&
+
\frac{2\pi}{N}\,
\rho_{h}
\left(
\kappa^{\,}_{2}(g)\,\kappa^{\,}_{2}(h)
+
\kappa^{\,}_{1}(g)\,\kappa^{\,}_{2}(h)
\right).
\end{align}
\end{subequations}
Using the identity \eqref{eq:useful cocycle identity} delivers
\begin{align}
\nu^{\,}_{1\wedge 2}(g,h)=
\nu^{\,}_{1}(g,h)
+
\nu^{\,}_{2}(g,h)
+
\frac{2\pi}{N}\,
\rho_{h}\,
\kappa^{\,}_{1}(g)\kappa^{\,}_{2}(h).
\end{align}
We write the stacked indices compactly as
\begin{subequations}
\label{eq:stacked indices oe}
\begin{align}
&
\kappa^{\,}_{1\wedge2} = \kappa^{\,}_{1} - \kappa^{\,}_{2},
\\
&
\nu^{\,}_{1\wedge2}=
\nu^{\,}_{1}
+
\nu^{\,}_{2}
+
\frac{2\pi}{N}\,
\kappa^{\,}_{1}\cup\kappa^{\,}_{2}.
\end{align}
\end{subequations}
Here too, the non-commutativity
(\ref{eq:non-commutative cup product})
holds.

\subsection{Summary of the stacking rules}

We considered the stacking of two IPT phases respectively characterized by the classifying indices $(\mu_1, [(\kappa_1,\nu_1)])$ and $(\mu_2, [(\kappa_2,\nu_2)])$ by distinguishing four cases according to $(\mu_1, \mu_2) \in \{0,1\}^2$.
We obtained the stacking rules \eqref{eq:stacked indices ee}, \eqref{eq:stacked indices oo},
\eqref{eq:stacked indices eo}, and \eqref{eq:stacked indices oe}.
These can be combined into a single stacking rule:
\begin{align}
\label{eq:stacking rules combined}
\begin{split}
&
\mu^{\,}_{1\wedge2}=
\mu^{\,}_{1}+\mu^{\,}_{2} \mod 2,
\\
&
\kappa^{\,}_{1\wedge2}=
\kappa^{\,}_{1} + (-1)^{\mu^{\,}_{1}}\,\kappa^{\,}_{2},
\\
&
\nu^{\,}_{1\wedge2}=
\nu^{\,}_{1} +\nu^{\,}_{2}
+
(-1)^{\mu^{\,}_{1}+1}\,
\frac{2\pi}{N}\kappa^{\,}_{1}\cup\kappa^{\,}_{2}.
\end{split}
\end{align}
One can identify from these formulae two sources of non-Abelian stacking: non-trivial $\mu_1$ and the (graded) anti-symmetry of the cup product.

Since the index $\mu=0$ identifies symmetry protected topological (SPT) phases, we can write the general stacking group structure of invertible parafermionic topological (IPT) phases as
\begin{equation}
\text{Inv}^{\mathrm{pf}}_N(G) \,\cong\, 
\text{SPT}^{\mathrm{pf}}_N(G) \,\rtimes\, 
\mathbb{Z}_2  
\end{equation}
The semi-direct product captures the non-Abelian stacking due to non-trivial $\mu_1$.
In stark contrast to bosons and fermions, parafermionic SPTs can themselves realize non-Abelian stacking. 
This is a consequence of the (graded) anti-symmetry of the cup-product, which we demonstrate explicitly in the example of Sec.\ \ref{sec:stacking example}.

\begin{figure}[t]
\centering
\includegraphics[width=1.0\linewidth]{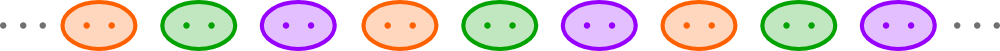}
\caption{\label{fig:hamiltonian-basepace}
The $3L$ sites of a chain are depicted as ellipses
colored in orange, green, and purple. Each ellipse contains two dots.
The left (right) dot of an orange ellipse represents the parafermion
$\xi^{\,}_{3j+1}$ ($\eta^{\,}_{3j+1}$).
The left (right) dot of a green ellipse represents the parafermion
$\xi^{\,}_{3j+2}$ ($\eta^{\,}_{3j+2}$).
The left (right) dot of a purple ellipse represents the parafermion
$\xi^{\,}_{3j+3}$ ($\eta^{\,}_{3j+3}$).
This rule fixes the ordering of the parafermions and the Hilbert space.
}
\end{figure}

\section{IPT phases with 
$\Z^{\,}_N\times\Z^{\,}_N\times\Z^{\mathrm{pf}}_N$ symmetry}
\label{sec:stacking example}

Let $N$ be an odd prime-power and assume that the
protecting symmetry group is given by the direct product
\begin{equation}  
\mathcal{G}\cong
\Z^{\mathrm{pf}}_{N}\times G^{\,}_{b},
\qquad
G^{\,}_{b}:=
\left(\Z^{\,}_{N}\times\Z^{\,}_{N}\right).
\label{eq:def-Gpf-for-explicit-example}
\end{equation}
We are going to introduce a family of local parafermionic Hamiltonians
with the symmetry group
(\ref{eq:def-Gpf-for-explicit-example})
such that their IPT phases are characterized by the triplets%
~\footnote{ 
Equation (\ref{eq: triplets rule example})
follows from Eqs.\
(\ref{eq:triplet-projective-indices-even-case})
and
(\ref{eq:triplet-projective-indices-odd-case})
applied to the symmetry group
(\ref{eq:def-Gpf-for-explicit-example}).
The decomposition into
$\mu=0,1$,
$(\kappa^{\,}_{1},\kappa^{\,}_{2})\in\{0,1,\ldots,N\!-\! 1\}^{2}$
and
$\nu\in\{0,1,\ldots,N \!-\! 1\}$
is a consequence of $\mathcal{G}$ being a trivial group extension of
$G^{\,}_{b}$ by $\Z^{\mathrm{pf}}_{N}$ on the one hand,
and the fact that $G^{\,}_{b}$ is a direct product of
$\mathbb{Z}^{\,}_{N}$ with itself on the other hand.   
}
\begin{equation}
[\mu,(\kappa^{\,}_{1},\kappa^{\,}_{2}),\nu]\in
\{0,1\}\times
\{0,1,\ldots,N\!-\! 1\}^{2}\times
\{0,1,\ldots,N \!-\! 1\}
\label{eq: triplets rule example}
\end{equation}
that encode their boundary projective representation
and obey the stacking rules
\begin{equation}
\begin{split}
&
[\mu,\, (\kappa^{\,}_{1}, \kappa^{\,}_{2}),\, \nu]
\, \wedge \,
[\tilde{\mu},\, (\tilde{\kappa}_{1}, \tilde{\kappa}_{2}),\, \tilde{\nu}]=
\\
&\qquad
\left[
\mu
+
\tilde{\mu}, \,\,
\kappa
+
(-1)^{\mu}\,
\tilde{\kappa}, \,\,
\nu
+
\tilde{\nu}
-
(-1)^{\mu}\,
\left(
\kappa^{\,}_{1}\,\tilde{\kappa}_{2}
-
\kappa^{\,}_{2}\,\tilde{\kappa}_{1}
\right)
\right].
\end{split}
\label{eq: stacking rule example}
\end{equation}
Remarkably, these stacking rules are the composition rules of the
semi-direct product
\begin{subequations}
\label{eq:relation-to-Heisenberg-group}
\begin{equation}
\mathrm{Heis}(\Z^{\,}_{N})\rtimes\Z^{\,}_{2},
\label{eq:relation-to-Heisenberg-group-a}
\end{equation}
where $\mathrm{Heis}(\Z^{\,}_{N})$ is the Heisenberg group (i.e., the
group of 3-by-3 unitriangular matrices) over the integer modulo ring
$\Z^{\,}_{N}$,
\begin{equation}
\mathrm{Heis}(\mathbb{Z}^{\,}_{N})=
\left\{
\begin{pmatrix}
1&a&z\\
0&1&b\\
0&0&1
\end{pmatrix}
\Big\vert
a,b,z\in\mathbb{Z}^{\,}_{N}
\right\}.
\label{eq:relation-to-Heisenberg-group-b}
\end{equation}
\end{subequations}
The group (\ref{eq:relation-to-Heisenberg-group-a})
is non-Abelian both because of the semi-direct product and because 
$\mathrm{Heis}(\mathbb{Z}^{\,}_{N})$ has non-trivial commutators.

We first construct the parafermionic Hilbert space as is depicted in
Fig.\ \ref{fig:hamiltonian-basepace}. We consider a chain made
of $3L$ sites.
These are represented by ellipses alternatingly  colored in orange, green, and purple.
Each site carries a pair of order-$N$ parafermions $(\xi, 
\eta)$, which are denoted by a left ($\xi$) and right ($\eta$) dot within each ellipse.
Orange, green, and purple ellipses are labeled $3j+1$, $3j+2$, and $3j+3$, respectively. 
The ordered set of parafermions for the chain is then
\begin{equation}
\left\{
\xi_{3j+1},\eta_{3j+1},\,
\xi_{3j+2},\eta_{3j+2},\,
\xi_{3j+3},\eta_{3j+3}
\right\}^{L-1}_{j=0}.
\label{eq:ordered-parafermion-6L-sites}
\end{equation}

\begin{figure}[t]
\centering
\includegraphics[width=1.0\linewidth]{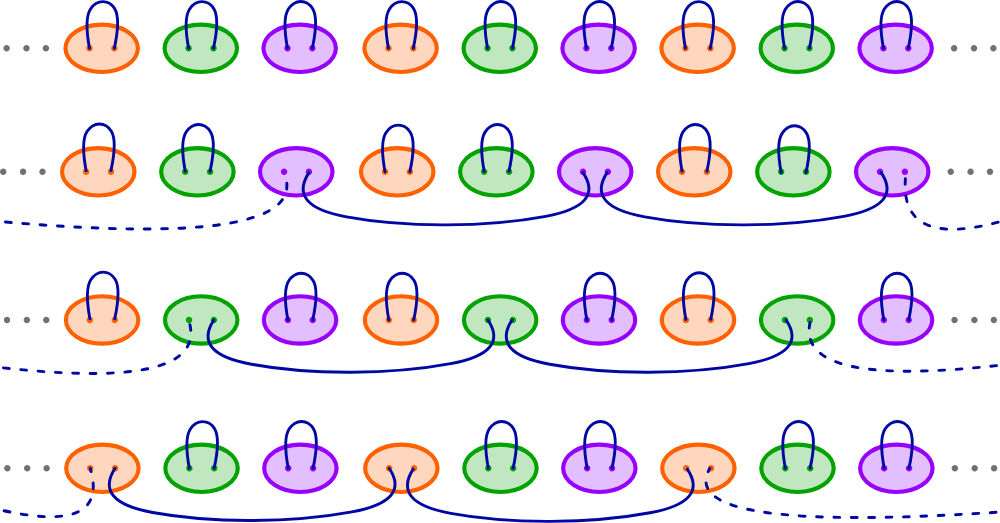}
\caption{\label{fig:hamiltonian-all}
The family of Hamiltonians
(\ref{eqs: generating hamiltonians})
are quadratic forms in the parafermionic operators.
Sites of the chain (ellipses) are assigned pairs of parafermionic operators (dots) as explained
in Fig.\ \ref{fig:hamiltonian-basepace}.
Two dots connected by a bond represent the product of the parafermion operators at its endpoints, with the operator at the left endpoint multiplying the operator at the right endpoint from the left. }
\end{figure}

Define the four Hamiltonians 
\small
\begin{subequations}
\label{eqs: generating hamiltonians}
\begin{align}
&
H^{(\mathsf{b})}_{[0, (0,0), 0]}
:=
\nonumber\\
&
-
\lambda\!\!
\sum_{j=0}^{L-1-\mathsf{b}}\!
\bigl(
\xi_{3j+1}^{\phantom{\dag}}\eta_{3j+1}^{\dag}
+
\xi_{3j+2}^{\phantom{\dag}}\eta_{3j+2}^{\dag}
+
\xi_{3j+3}^{\phantom{\dag}}\eta_{3j+3}^{\dag}
+\mathrm{H.c.}
\bigr),
\label{eq: [0,(0,0),0] Ham.}
\\
&
H^{(\mathsf{b})}_{[1, (0,0), 0]}:=
\nonumber\\
&
-
\lambda\!\!
\sum_{j=0}^{L-1-\mathsf{b}}\!
\bigl(
\xi_{3j+1}^{\phantom{\dag}}\eta_{3j+1}^{\dag}
+
\xi_{3j+2}^{\phantom{\dag}}\eta_{3j+2}^{\dag}
+
\underline{\eta_{3j+3}^{\phantom{\dag}}\xi_{3(j+1)+3}^{\dag}}
+\mathrm{H.c.}
\bigr),
\label{eq: [1,(0,0),0] Ham.}
\\
&
H^{(\mathsf{b})}_{[1, (0,1), 0]}:=
\nonumber\\
&
-
\lambda\!\!
\sum_{j=0}^{L-1-\mathsf{b}}\!
\bigl(
\xi_{3j+1}^{\phantom{\dag}}\eta_{3j+1}^{\dag}
+
\underline{\eta_{3j+2}^{\phantom{\dag}}\xi_{3(j+1)+2}^{\dag}}
+
\xi_{3j+3}^{\phantom{\dag}}\eta_{3j+3}^{\dag}
+\mathrm{H.c.}
\bigr),
\label{eq: [1,(0,1),0] Ham.}
\\
&
H^{(\mathsf{b})}_{[1, (1,0), 0]}:=
\nonumber\\
&
-
\lambda\!\!
\sum_{j=0}^{L-1-\mathsf{b}}\!
\bigl(
\underline{\eta_{3j+1}^{\phantom{\dag}}\xi_{3(j+1)+1}^{\dag}}
+
\xi_{3j+2}^{\phantom{\dag}}\eta_{3j+2}^{\dag}
+
\xi_{3j+3}^{\phantom{\dag}}\eta_{3j+3}^{\dag}
+\mathrm{H.c.}
\bigr).
\label{eq: [1,(1,0),0] Ham.}
\end{align}
\end{subequations}
\normalsize
The superscript $\mathsf{b}=0,1$
implements periodic and open boundary conditions, respectively.
The subscript denotes the classifying indices 
associated to the corresponding IPT phases, as we will be shown explicitly.
The coupling $\lambda>0$ has units of energy.
The summands
are grouped according to flavor.
These four Hamiltonians are depicted schematically in Fig.\
\ref{fig:hamiltonian-all}.

Each Hamiltonian is a sum of pairwise commuting terms and has a
non-degenerate ground state under periodic boundary conditions
($\mathsf{b}=0$).  Changing the boundary condition leaves
$H^{(\mathsf{b})}_{[0, (0,0), 0]}$ unchanged.  For
$H^{(\mathsf{b})}_{[1, (0,0), 0]}$,
$H^{(\mathsf{b})}_{[1, (0,1), 0]}$,
and
$H^{(\mathsf{b})}_{[1, (1,0), 0]}$,
choosing open boundary conditions
$\mathsf{b}=1$ leads to the respective terms
$\eta_{3L}^{\phantom{\dag}}\, \xi_{3}^{\dag}$,
$\eta_{3L-1}^{\phantom{\dag}}\, \xi_{2}^{\dag}$, and
$\eta_{3L-2}^{\phantom{\dag}}\, \xi_{1}^{\dag}$ dropping out
of the sums.  These imply an $N$-fold ground-state degeneracy under
open boundary conditions.  We construct from the local charges
$Q^{\,}_{j}$ the operators
\begin{align}
U^{\,}_{k}:=
\prod_{j=0}^{3L-1}
Q_{3j+k},
\qquad
k=1,2,3.
\end{align}
These are of order $N$ and commute pairwise.
One may think of these as the global charge restricted to each
parafermion flavor. They commute with the above
Hamiltonians and we fix the protecting symmetry group to be
\begin{equation}
\mathcal{G}=
\langle
Q,U^{\,}_{1},U^{\,}_{2} 
\rangle\cong
\Z^{\mathrm{pf}}_{N}\times(\Z^{\,}_{N}\times\Z^{\,}_{N}),
\end{equation}
as announced at the beginning of this section.  Thus, the Hamiltonians
in Eq.\ \eqref{eqs: generating hamiltonians} are indeed
representatives of $\mathcal{G}$-symmetry protected topological
phases.

We now compute the classifying indices of the Hamiltonians and show
that they are correctly labeled.  Since the ground state of
$H^{(\mathsf{b})}_{[0,(0,0),0]}$ remains non-degenerate under
open boundary conditions $\mathsf{b}=1,$ it is clear that it
corresponds to the trivial phase classified by the indices
$[0,(0,0),0]$.  For the other Hamiltonians, the single summand dropping out when choosing open boundary conditions causes the
boundary algebra to be generated by an odd number of parafermions.
This implies that the first classifying index is $\mu=1$ for the
three Hamiltonians
\eqref{eq: [1,(0,0),0] Ham.}-\eqref{eq: [1,(1,0),0] Ham.}.
Using an auxiliary right-infinite parafermion
$\gamma^{\,}_{\infty}$, the boundary parafermion algebras for
the respective Hamiltonians are
\begin{subequations}
\begin{align}
&
\mathfrak{P}^{\,}_{[1,(0,0),0]}=
\langle
\xi^{\,}_{1},
\eta^{\,}_{1},
\eta^{\,}_{2},
\xi^{\,}_{2},
\underline{\xi^{\,}_3},
\gamma^{\,}_{\infty}
\rangle,
\\
&
\mathfrak{P}^{\,}_{[1,(0,1),0]}=
\langle
\xi^{\,}_{1},
\eta^{\,}_{1},
\underline{\xi^{\,}_{2}},
\gamma^{\,}_{\infty}
\rangle,
\\
&
\mathfrak{P}^{\,}_{[1,(1,0),0]}=
\langle
\underline{\xi^{\,}_{1}},
\xi^{\,}_{2},
\eta^{\,}_{2},
\gamma^{\,}_{\infty}
\rangle,
\end{align}
\end{subequations}
where we underlined the parafermion stemming from the dropped
summation terms. The boundary algebra $\mathfrak{P}^{\,}_{[1,(1,0),0]}$
was expanded with the neighboring parafermion pair
$\xi^{\,}_{2},
\eta^{\,}_{2}$
to avoid the degenerate case of vanishing boundary
parafermion charge. We now consider the action of the global
symmetries
$Q, U^{\,}_{1},$
and
$U^{\,}_{2}$
on these boundary degrees of freedom. For the Hamiltonian
$H^{(\mathsf{b})}_{[1,(0,0),0]}$,
they produce for each boundary parafermion (excluding the auxiliary
$\gamma^{\,}_{\infty}$)
the following phases,
\begin{subequations}
\begin{align}
&
Q:\quad (\w^{\,}, \w^{\,}, \w^{\,}, \w^{\,}, \w^{\,}),
\\
&
U^{\,}_{1}:\quad (\w^{\,}, \w^{\,}, 1, 1, 1),
\\
&
U^{\,}_{2}:\quad (1, 1, \w^{\,}, \w^{\,}, 1),
\end{align}
\end{subequations}
Analogously, for the Hamiltonian $H^{(\mathsf{b})}_{[1,(0,1),0]}$
\begin{subequations}
\begin{align}
&
Q:\quad (\w^{\,}, \w^{\,}, \w^{\,}),
\\
&
U^{\,}_{1}:\quad (\w^{\,}, \w^{\,}, 1),
\\
&
U^{\,}_{2}:\quad (1, 1, \w^{\,}),
\end{align}
\end{subequations}
and the Hamiltonian $H^{(\mathsf{b})}_{[1,(1,0),0]}$
\begin{subequations}
\begin{align}
&
Q:\quad (\w^{\,}, \w^{\,}, \w^{\,}),
\\
&
U^{\,}_{1}:\quad (\w^{\,}, 1, 1),
\\
&
U^{\,}_{2}:\quad (1, \w^{\,}, \w^{\,}).
\end{align}
\end{subequations}
We must now find boundary projections
$Q^{(\mathsf{b})}$,
$U^{(\mathsf{b})}_{1}$,
and
$U^{(\mathsf{b})}_{2}$
in terms of the boundary parafermions
whose action on the degrees of freedom match the global ones recorded above.
One may verify that such representations are given:\\
for
$H^{(\mathsf{b})}_{[1, (0,0),0]}$ by
\begin{subequations}
\label{eqs: [1, (0,0), 0] bdd. rep}
\begin{align}
&
Q^{(\mathsf{b})}\equiv
Q^{\,}_{1}\,
Q^{\,}_{2}\,
\left(\w^{\frac{N+1}{2}}\,\xi^{\,}_{3}\,\gamma^{\dag}_{\infty}\right),
\\
&
U^{(\mathsf{b})}_{1}\equiv
Q^{\,}_{1},
\\
&
U^{(\mathsf{b})}_{2}\equiv
Q_{2},
\end{align}
\end{subequations}
for $H^{(\mathsf{b})}_{[1 (0,1),0]}$ by
\begin{subequations}
\label{eqs: [1, (0,1), 0] bdd. rep}
\begin{align}
&
Q^{(\mathsf{b})}\equiv
Q^{\,}_{1}\,
\left(\w^{\frac{N+1}{2}}\,\xi^{\,}_{2}\,\gamma^{\dag}_{\infty}\right),
\\
&
U^{(\mathsf{b})}_{1}\equiv
Q^{\,}_{1},
\\
&
U^{(\mathsf{b})}_{2}\equiv
\left(\w^{\frac{N+1}{2}}\,\xi^{\,}_{2}\,\gamma^{\dag}_{\infty}\right),
\end{align}
\end{subequations}
and for $H^{(\mathsf{b})}_{[1, (1,0),0]}$ by
\begin{subequations}
\label{eqs: [1, (1,0), 0] bdd. rep}
\begin{align}
&
Q^{(\mathsf{b})}\equiv
\left(\w^{\frac{N+1}{2}}\,\xi^{\,}_{1}\,\xi^{\dag}_{2}\right)
\left(\w^{\frac{N+1}{2}}\,\eta^{\,}_{2}\,\gamma^{\dag}_{\infty}\right),
\\
&
U^{(\mathsf{b})}_{1}\equiv
\w^{\frac{N+1}{2}}\,
\xi^{\dag}_{1}\,
\gamma^{\dag}_{\infty},
\\
&
U^{(\mathsf{b})}_{2}\equiv
Q^{\,}_{2}.
\end{align}
\end{subequations}
Moreover, all these operators commute with their respective Hamiltonians.

We now use these boundary representations to compute the classifying
indices.  As mentioned, the first index is always $\mu=1$ as there is
an odd number of boundary parafermions. The second index
$\kappa=(\kappa^{\,}_{1},\kappa^{\,}_{2})$ is an element of
\begin{equation}
H^{1}[\Z^{\,}_{N}\times \Z^{\,}_{N},\, \Z^{\,}_{N}]\cong
\Z^{\,}_{N} \times \Z^{\,}_{N}.
\end{equation}
It corresponds to the order of the
Hermitian conjugated auxiliary parafermion
$\gamma^{\dag}_{\infty}$ in the boundary representations
$U^{(\mathsf{b})}_{1}$ ($\kappa^{\,}_{1}$)
and
$U^{(\mathsf{b})}_{2}$ ($\kappa^{\,}_{2}$).
From definitions
\eqref{eqs: [1, (0,0), 0] bdd. rep},
\eqref{eqs: [1, (0,1), 0] bdd. rep},
and
\eqref{eqs: [1, (1,0), 0] bdd. rep}
one deduces
\begin{equation}
(\kappa^{\,}_{1},\kappa^{\,}_{2})=
(0,0),\
(0,1),\
(1,0),
\end{equation}
respectively.
The third index $\nu$ is an element of
\begin{equation}
H^{2}[\Z^{\,}_{N}\times \Z^{\,}_{N},\,\Z^{\,}_{N}]\cong
\Z^{\,}_{N}.
\end{equation}
It measures the projective equivalency class of the
boundary representation of the internal symmetry
$\Z^{\,}_{N}\times\Z^{\,}_{N}$,
which may be computed by considering the commutation
relation between
$U^{(\mathsf{b})}_{1}$ and $U^{(\mathsf{b})}_{2}$.
In fact, in the three cases above they always commute, whence $\nu=0$.
We have thus verified that
the classifying indices match the subscripts of Hamiltonians
\eqref{eqs: generating hamiltonians}.

Recall that $N$ is odd.  To find the explicit stacking rule
(\ref{eq: stacking rule example}) from Eq.\ 
\eqref{eq:stacking rules combined},
one must compute the group cohomologies and write out the
cup product.
The latter follows from the fact that the non-degenerate alternating
bilinear form over two-dimensional spaces is unique up to scalar factors, which may be viewed as
the determinant.
Since the cup product of two $1$-cocycles is anti-symmetric, 
we can conclude
\begin{equation}
\kappa\cup\tilde{\kappa}=
\kappa^{\,}_{1}\,\tilde{\kappa}_{2}
-
{\kappa}_{2}\,\tilde{\kappa}_{1}.
\end{equation}

We label the three non-trivial IPT phases by
\begin{subequations}
\begin{align}
&
S:=[1,\, (0,0),\, 0],
\label{eq: S phase}
\\
&
\widetilde{X}:=[1,\, (1,0),\, 0],
\label{eq: tildeX phase}
\\
&
\widetilde{Y}=[1,\, (0,1),\, 0].
\label{eq: tildeY phase}
\end{align}
Using our stacking rules, we see that the phase $S$ may be used to
flip the first classifying index, giving us the SPT phases,
and we define
\begin{align}
X &:= [0,\, (1,0),\, 0] = \widetilde{X} \wedge S,
\\
Y &:= [0,\, (0,1),\, 0] = \widetilde{Y} \wedge S.
\end{align}
In particular, applying the stacking-group commutator yields the phase
\begin{align}
Z:=
\bigl[ X, Y \bigr]^{\varphi(N)}=
[0,\, (0,0),\, -2]^{\varphi(N)}=
[0,\, (0,0),\, 1],
\end{align}
\end{subequations}
where $\varphi$ is Euler's totient function and we used the fact that
$N$ is odd.  It is thus clear that the phases $\widetilde{X},
\widetilde{Y}$ in combinations with the phase $S$ generate all other
\mbox{$\mathcal{G}\cong\Z^{\mathrm{pf}}_{N}\times(\Z^{\,}_{N}\times\Z^{\,}_{N})$%
-protected topological phases}.
Moreover, the isomorphism of the SPT phases
(the $\mu=0$ stacking subgroup)
to the Heisenberg group
$\mathrm{Heis}(\mathbb{Z}^{\,}_{N})$
is given explicitly by the mapping
\begin{equation}
[0,\, (\kappa^{\,}_{1}, \kappa^{\,}_{2}),\, \nu]
\, \longmapsto \,
\begin{pmatrix}
1 & \kappa^{\,}_{1} & \frac{\kappa^{\,}_{1} \kappa^{\,}_{2} - \nu}{2}\\
0 & 1 & \kappa^{\,}_{2}\\
0 & 0 &1
\end{pmatrix},
\end{equation}
which is well-defined due to $N$ being odd.
Finally, stacking the SPT phases from the right with $S$ leads to the semi-direct product in the total stacking group,
\begin{equation}
\mathrm{Heis}(\Z^{\,}_{N})\rtimes\Z^{\,}_{2}.
\end{equation}

\section{Conclusion}
\label{sec:conclusion}

In this work, we have studied the
invertible parafermionic topological (IPT) phases
in one-dimensional space. Building on Ref.\ \cite{Aksoy2022}, 
we introduced a triplet of indices $(\mu,[(\kappa,\nu)])$
that characterizes projective representations of global
symmetry $\mathcal{G}$ acting on order-$N$ parafermions.
These indices in turn classify distinct one-dimensional
IPT phases of matter with symmetry $\mathcal{G}$.
While we have only considered unitary symmetries in this work,
our results can be extended to anti-unitary symmetries such as
time-reversal, see for example Refs.\ \cite{Turzillo2019,Aksoy2022}.

We have then derived the so-called stacking rules
that prescribes how two IPT phases can be merged to 
produce a third one. Stacking rules can be thought of
as a composition rule that gives the IPT phases a group structure.
We have shown that as opposed the invertible topological 
phases of bosons and fermions, IPT phases can form a non-Abelian group
under stacking. 

In Sec.\ \ref{sec:stacking example} we 
explicitly constructed Hamiltonians that realize IPT phases
of order odd prime power $N$ 
with $\Z^{\,}_N\times\Z^{\,}_N\times\Z^{\mathrm{pf}}_N$ symmetry.
Therein, we showed that under stacking the IPT phases form the 
non-Abelian group $\mathrm{Heis}(\Z_N)\rtimes\Z_2$ of order $2N^3$,
where $\mathrm{Heis}(\Z_N)$ is the Heisenberg group of order $N^3$. 

In Sec.\ \ref{subsec:general N}, we discussed the classification
of IPT phases for general integer $N$. Given the prime decomposition
of $N$, we have shown that the degrees of freedom split into 
flavors, one for each prime factor. When the global symmetries
factorizes on each flavor as in Eq.\ 
\eqref{eq:factorization holds for any symmetry}, 
the classification simplifies and is given in Eq.\ 
\eqref{eq:general N collection of indices}. A natural question 
to ask would be if the classification in terms of indices 
\eqref{eq:general N collection of indices} also holds
in the more general 
case when Eq.\ \eqref{eq:factorization holds for any symmetry} 
does not hold, which we leave for future work. 

One can relate the IPT phases of order-$N$ parafermions 
to bosonic gapped phases realized by $\Z_N$ clock and shift 
variables~\cite{Fradkin1980,Fendley2012,Alicea2016}. While, parafermionic
SPT phases are generally mapped to bosonic SPT phases, the IPT phases that 
are not SPTs are mapped to spontaenous symmetry breaking phases. 
It would be interesting to match the stacking rules 
obeyed by IPT phases
with the maps between bosonic gapped phases.

\section*{Acknowledgments}
We thank Christopher Mudry and Claudio Chamon
for discussions and collaboration in the early stages of this work.
We thank Christopher Mudry for comments on the manuscript and and for identifying an issue in Sec.\ \ref{subsec:general N}.
A.L.G.M. acknowledges support from 
the Swiss Study Foundation and the Swiss 
State Secretariat for Education, Research and Innovation.
\"O.M.A. was supported in parts 
by SNSF Grant No.~214429,
by NSF Grant No.~DMR2022428,
and by the 
Simons Collaboration on Ultra-Quantum Matter, 
which is a grant from the Simons Foundation (651444, SHS).

\vfill

\appendix
\addappheadtotoc
\onecolumngrid

\section{Group cohomology}
\label{appsec:Group Cohomology}

\noindent Given two groups $G$ and $M$, an \textit{$n$-cochain} is the map
\begin{equation}
\begin{split}
\phi \colon G^{n} \to&\, M,
\\
(g^{\,}_{1},g^{\,}_{2},\cdots, g^{\,}_{n})\mapsto&\,
\phi(g^{\,}_{1},g^{\,}_{2},\cdots, g^{\,}_{n}),
\end{split}
\end{equation}
that maps an $n$-tuple $(g^{\,}_{1},g^{\,}_{2},\cdots, g^{\,}_{n})$
to an element $\phi(g^{\,}_{1},g^{\,}_{2},\dots,g^{\,}_{n})\in M$.
The set of all $n$-cochains from $G^{n}$ to $M$
is denoted by $C^{n}(G,M)$.
We define an $M$-valued 0-cochain to be an element of the group $M$
itself, i.e., $C^{0}(G,M)= M$.
Henceforth, we will denote
the group composition rule in $G$ by $\cdot$
and the group composition rule in $M$ additively by $+$
($-$ denoting the inverse element). 
For our purposes,
the Abelian group $M$ will be either $U(1)$ or the cyclic 
group $\Z_N$.

In the latter case, we allow a non-trivial group action of $G$ on $M = \Z_N$.
Given a group homomorphism $\rho_g\colon G \to \Z_N$,
we define the group action
\begin{equation}
\begin{split}
\rho_g
\colon
\Z_N \to& \Z_N,
\\
m\ \mapsto&\
\rho_g\,m,
\end{split}
\end{equation}
which is just multiplication by integers modulo $N$.
In our setting, this group action will always be involutive, 
i.e., ${\rho^2_g=1\text{ mod }N}$.
The map $\rho$ indicates whether an element
$g\in G$ implements a (special) charge conjugation or not.
Note that since $\rho_g$ is a group homomorphism one has
$\rho_{g\cdot h}=\rho_g\,\rho_h=\rho_h\,\rho_g$. 
When $M=U(1)$, we view $G$ as acting with the trivial group action, which corresponds to our restriction to unitary representations of symmetries.

We define the map $\delta^{n}_{\rho}$
\begin{subequations}\label{appeq:definition delta n}
\begin{equation}
\begin{split}
\delta^{n}_{\rho}\colon
C^{n}(G,M)\ \to&\ C^{n+1}(G,M),
\\
\phi\ \mapsto&\ \left(\delta^{n}_{\rho}\phi\right),
\end{split}
\end{equation}
from $n$-cochains to $(n+1)$-cochains such that
\begin{align}
\label{eq:def coboundary operator}
\left(\delta^{n}_{\rho}\phi\right)
(g^{\,}_{1},\cdots,g^{\,}_{n+1}):=
\phi(g^{\,}_{2},\cdots, g^{\,}_{n},g^{\,}_{n+1})
+
\sum_{i=1}^{n}
(-1)^{i}
\phi(g^{\,}_{1},\cdots,g^{\,}_{i}\cdot g^{\,}_{i+1},
\cdots,g^{\,}_{n+1})
-
(-1)^{n}\,
\rho_{g^{\,}_{n+1}}\,
\phi(g^{\,}_{1},\cdots,g^{\,}_{n}).
\end{align}
\end{subequations}
The map $\delta^{n}_{\rho}$
is called a \textit{coboundary operator}.

\textbf{Example $n=2$:}
The coboundary operator $\delta^{2}_{\rho}$ is defined by
\begin{align}
\left(\delta^{2}_{\rho}\phi\right)
(g^{\,}_{1},g^{\,}_{2},g^{\,}_{3})=&\,
\phi(g^{\,}_{2},g^{\,}_{3})
+
(-1)^{1}
\phi(g^{\,}_{1}\cdot g^{\,}_{2},g^{\,}_{3})
+
(-1)^{2}
\phi(g^{\,}_{1}, g^{\,}_{2}\cdot g^{\,}_{3})
-
(-1)^{2}\,\
\rho_{g^{\,}_{3}}\,
\phi(g^{\,}_{1},g^{\,}_{2})
\nonumber\\
=&\,
\phi(g^{\,}_{2},g^{\,}_{3})
-
\phi(g^{\,}_{1}\cdot g^{\,}_{2},g^{\,}_{3})
+
\phi(g^{\,}_{1}, g^{\,}_{2}\cdot g^{\,}_{3})
-
\rho_{g^{\,}_{3}}\,
\phi(g^{\,}_{1},g^{\,}_{2}).
\label{appeq:example delta2}
\end{align}

\textbf{Example $n=1$:}
The coboundary operator $\delta^{1}_{\rho}$ is defined by
\begin{align}
\left(\delta^{1}_{\rho}\phi\right)
(g^{\,}_{1},g^{\,}_{2})=&\,
\phi(g^{\,}_{2})
-
\phi(g^{\,}_{1}\cdot g^{\,}_{2})
+
\rho_{g^{\,}_{2}}\,
\phi(g^{\,}_{1}).
\label{appeq:example delta1}
\end{align}
One verifies the important identity
\begin{equation}
\Phi(g^{\,}_{1},g^{\,}_{2}):=
\left(\delta^{1}_{\rho}\phi\right)
(g^{\,}_{1},g^{\,}_{2})
\ \implies \
(\delta^{2}_{\rho}\Phi)(g^{\,}_{1},g^{\,}_{2},g^{\,}_{3})=0.
\label{appeq:one-cochain always obeys the cocycle conditions}
\end{equation}
Using the coboundary operator, we define two the sets
\begin{subequations}
\begin{align}
Z^{n}(G,M^{\,}_{\rho}):=
\mathrm{ker}(\delta^{n}_{\rho})=
\left\{
\phi \in C^{n}(G,M) \ |\
\delta^{n}_{\rho}\phi=0
\right\},
\end{align}
and
\begin{align}
B^{n}(G,M^{\,}_{\rho}):=
\mathrm{im}(\delta^{n-1}_{\rho})=
\left\{
\phi \in C^{n}(G,M) \ |\
\phi=\delta^{n-1}_{\rho}\phi',\,\,
\phi'\in C^{n-1}(G,M)
\right\}.
\end{align}
\end{subequations}
The cochains in $Z^{n}(G,M^{\,}_{\rho})$
are called \textit{$n$-cocycles}.
The cochains in $B^{n}(G,M^{\,}_{\rho})$
are called \textit{$n$-coboundaries}.
The action of the boundary operator on the elements of the group $M$
is sensitive to the homomorphism $\rho$.
For this reason, we label $M$ by
$\rho$ in $Z^{n}(G,M^{\,}_{\rho})$
and $B^{n}(G,M^{\,}_{\rho})$.
The importance of the coboundaries is that the identity
(\ref{appeq:one-cochain always obeys the cocycle conditions})
generalizes to
\begin{equation}
\phi=\delta^{n-1}_{\rho}\phi'
\ \implies\
\delta^{n}_{\rho}\phi=0.
\label{appeq:n-cochain always obeys the cocycle conditions}
\end{equation}
The $n$th cohomology group is defined as the quotient of the
$n$-cocycles by the $n$-coboundaries, i.e.,
\begin{align}
H^{n}(G,M^{\,}_{\rho}):=
Z^{n}(G,M^{\,}_{\rho})/B^{n}(G,M^{\,}_{\rho}).
\end{align}
The $n$th cohomology group $H^{n}(G,M^{\,}_{\rho})$
is an additive Abelian group.
We denote its elements by $[\phi]\in H^{n}(G,M^{\,}_{\rho})$,
i.e., the equivalence class
of the $n$-cocycle $\phi$.

Finally, we define the \textit{cup product} of a pair of $\Z_N$-valued cochains
as follows.
Given two cochains $\phi\in C^{n}(G,\Z_N)$ and $\theta\in C^{m}(G,\Z_N)$,
we produce the $(n+m)$-cochain
$(\phi \cup \theta)\in C^{n+m}(G,\Z_N)$
through
\begin{align}
\label{eq:def cup product}
(\phi \cup \theta)
(g^{\,}_{1},\cdots,g^{\,}_{n},g^{\,}_{n+1},\cdots,g^{\,}_{m})
:=
\rho_{g^{\,}_{n+1}\cdot g^{\,}_{n+2}\cdots g^{\,}_{n+m}}\,
\phi(g^{\,}_{1},\cdots,g^{\,}_{n})\,
\theta(g^{\,}_{n+1},\cdots,g^{\,}_{n+m})
\mod N,
\end{align}
where the multiplication of the cochains $\phi$ and $\theta$ is treated
as the multiplication of integers modulo $N$.
For instance, the cup product of a 1-cochain
$\alpha\in C^{1}\big(G,\Z_N\big)$
and a 2-cochain $\beta \in C^{2}\big(G,\Z_N\big)$ is
\begin{align}
(\alpha \cup \beta ) (g^{\,}_{1},g^{\,}_{2},g^{\,}_{3})=
\rho_{g^{\,}_{2}\cdot g^{\,}_3}\,
\alpha(g^{\,}_{1})\,
\beta(g^{\,}_{2},g^{\,}_{3}).
\end{align}
The cup product defined in Eq.\ \eqref{eq:def cup product} satisfies
\begin{align}
\delta^{n+m}_{\rho} (\phi \cup \theta)
=
\left( \delta^{n}_{\rho} \phi \cup \theta \right)
+
(-1)^{n}\,
\left( \phi \cup \delta^{m}_{\rho}\theta \right)
\label{eq:boundary of a cup product}
\end{align}
for two cochains $\phi\in C^{n}(G,M)$ and $\theta\in C^{m}(G,M)$.
Hence, the cup product of two cocycles is again a cocycle, as the
right-hand side of Eq.\ \eqref{eq:boundary of a cup product} vanishes in that case.

\section{Construction of the parafermionic symmetry group $\mathcal{G}$}
\label{appsec:construction of G}

As argued in Sec.\ \ref{subsec:PF Charge}, the IPT phases of
order-$N$ parafermions have a global $\Z^{\mathrm{pf}}_N$ parafermion
charge symmetry. 
In the presence of additional symmetries described by a group $G_b$, the global symmetry $\mathcal{G}$ 
is a group extension of $G_b$ by $\Z^{\mathrm{pf}}_N$, as we shall explain.

\subsection{Parafermion charge symmetry is a normal subgroup}
\label{appsubsec:Zn normal subgroup}

We demand that any element $g\in G_b$ realizes a locality-preserving 
transformation, i.e., it maps every operator to nearby operators. 
Let us consider two operators $A_j$ and $B_k$ with definite parafermion
charges $q_A$ and $q_B$, respectively, whose supports are 
spatially separated. They satisfy the algebra 
\eqref{eq:algebra well separated Pf ops}. Under the action  of 
$g\in G_b$, these operators are mapped to the operators $A'_j$
and $B'_k$ which, by $g$ being locality-preserving, must also have
supports that are spatially separated. In particular, they must
satisfy the same algebra \eqref{eq:algebra well separated Pf ops}, i.e.,
\begin{align}
\label{eq:well separated algebra in the app}
A'_j\, B'_k 
= 
\omega^{q_A\,q_B\mathrm{sgn}(k-j)}_N\,
B'_k\, A'_j.
\end{align}
We can write the operators $A'_j$ and $B'_k$ 
in terms of sums of operators with well-defined parafermion charges,
\begin{align}
\label{eq:expanding transformed separated ops}
A'_j = \sum_{\alpha=0}^{N-1} A'^{(\alpha)}_j,
\qquad
B'_k = \sum_{\beta=0}^{N-1} B'^{(\beta)}_k,
\end{align}
where $A'^{(\alpha)}_j$ ($B'^{(\beta)}_k$)
is an operator with definite parafermion charge $\alpha \,(\beta) \in \Z_N$.

On the one hand, the left-hand side of Eq.\
\eqref{eq:well separated algebra in the app} becomes
\begin{subequations}
\begin{align}
\begin{split}
A'_j\, B'_k 
&=
\sum_{\alpha,\beta=0}^{N-1}
A'^{(\alpha)}_j\,
B'^{(\beta)}_k
\\
&=
\sum_{\alpha,\beta=0}^{N-1}
\omega_N^{\alpha\,\beta\,\mathrm{sgn}(k-j)}
B'^{(\beta)}_k\,
A'^{(\alpha)}_j,    
\end{split}
\end{align}   
where we used the algebra \eqref{eq:algebra well separated Pf ops} for each definite parafermion-charge summand.
On the other hand, the right-hand side of Eq.\
\eqref{eq:well separated algebra in the app} 
becomes
\begin{align}
\omega^{q_A\,q_B\,\mathrm{sgn}(k-j)}_N\,
B'_k\, A'_j
=
\omega^{q_A\,q_B\,\mathrm{sgn}(k-j)}_N\,
\sum_{\alpha,\beta=0}^{N-1}
B'^{(\beta)}_k\,
A'^{(\alpha)}_j.
\end{align}  
\end{subequations}
Matching the two sides of the equation and comparing coefficients
implies that only the operators with charges satisfying
$\alpha\,\beta = q_A\,q_B \text{ mod }N$ can appear in the decompositions
\eqref{eq:expanding transformed separated ops}. By choosing
$q_A=q_B=1$ we find 
\begin{align}
\alpha\,\beta = 1 \mod{N}.
\end{align}
For a fixed $\alpha$ this equation is satisfied by at most a single 
$\beta$. 
Hence, the sums in \eqref{eq:expanding transformed separated ops} reduce to a single term associated to a parafermion charge $\alpha$ and $\beta$.
In other words, locality dictates
that any global internal symmetry operator must map an operator
with a definite parafermion charge to another operator with a definite
parafermion charge, i.e., the $\Z^{\mathrm{pf}}_N$ grading must be 
preserved.

A consequence of the fact that every symmetry preserves the
$\Z^{\mathrm{pf}}_N$ grading is that $\Z^{\mathrm{pf}}_N$ must be 
a normal subgroup of $\mathcal{G}$, since its block-diagonal form must
be preserved under conjugation by the 
symmetries up to permutations of blocks. 
To be precise, a $\Z^{\mathrm{pf}}_N$-eigensector 
with definite parafermion charge $q$ can only be mapped
to the eigensector 
\begin{align}
g :\, q \mapsto \rho_g\,q + \kappa,
\end{align}
for any element $g\in G_b$. Here, $\rho_g$ corresponds
to the possible action of $g$ on $\mathbb{Z}^{\mathrm{pf}}_N$
while the $\kappa=0,1,\cdots,N-1$ denotes the parafermion charge
carried by the representation of $g$. 

\subsection{Parafermionic symmetry group $\mathcal{G}$}
\label{appsubsec:G as an extension}

We are now equipped to describe the structure 
of the total symmetry group $\mathcal{G}$. 
Since $\Z^{\mathrm{pf}}_N\subset \mathcal{G}$ is 
a normal subgroup, $\mathcal{G}$ is given by a 
group extension of $G_b\cong G/\Z^{\mathrm{pf}}_N$. 
We denote by $\tilde{g}$ the elements of group $\mathcal{G}$. 
Each element $\tilde{g}$ can also be alternatively written
as the pair $(z,g)$ where $z\in \Z^{\mathrm{pf}}_N$
and $g\in G_b$. We denote by $\circ$ the group composition
rule in $\mathcal{G}$ while we use the addition $+$ and 
multiplication $\cdot$ for group composition rules in 
$\Z^{\mathrm{pf}}_N$ and $\mathcal{G}$. 

The group extension is specified by the short-exact sequence
\begin{align}
1\to
\Z^{\mathrm{pf}}_{N}\underset{\iota}{\to}
\mathcal{G}\underset{\pi}{ \to}
G^{\,}_{b} \to 1,
\end{align}
where $\iota$ is an injective inclusion map and $\pi$ a surjective
quotient map such that ${\mathrm{im}\,\iota=\mathrm{ker}\,\pi}$.
The map $\iota$ describes how elements of $\Z^{\mathrm{pf}}_N$
are embedded in the group $\mathcal{G}$, respectively. 
The map $\pi$ specifies the quotient 
\begin{align}
\begin{split}
G_b 
&\cong 
\mathcal{G}/\mathrm{ker}\,\pi\\
& = \mathcal{G}/\mathrm{im}\,\iota\\
& \cong \mathcal{G}/\Z^{\mathrm{pf}}_N.
\end{split}
\end{align}
Since $\Z^{\mathrm{pf}}_N$ is a normal subgroup the quotient 
is a group. We define two more maps 
\begin{subequations}
\begin{align}
s: G_b \to \mathcal{G}, \qquad \tau: \mathcal{G}\to \Z^{\mathrm{pf}}_N,
\end{align}
such that  
\begin{align}
\pi(s) = \mathrm{id}_{G_b}, 
\quad
\tau(\iota) = \mathrm{id}_{\Z^{\mathrm{pf}}_N},
\quad
\tau(\iota(z)\circ \tilde{g}) = z\,\tau(\tilde{g}),
\quad
\tau(s(g)) = 0,
\end{align}
\end{subequations}
where $\mathrm{id}_{\mathcal{G}}$ 
and $\mathrm{id}_{\Z^{\mathrm{pf}}_N}$
are identity isomorphisms on the groups $G_b$
and $\Z^{\mathrm{pf}}_N$. 
The map $s$ is injective but \textit{not necessarily} a homomorphism
and called a section of the projection $\pi$. 
There is always a choice of a section such that $\tau(s(g)) = 0$,
which we make.
The map $\tau$
is surjective and called the \textit{trivialization}.

There exist a bijection between 
the groups $\mathcal{G}$ and $\Z^{\mathrm{pf}}_N\times G_b$ 
as \textit{sets} but not necessarily as groups. 
To see this, we note that any element 
$\tilde{g}\in\mathcal{G}$
is mapped to a pair $(z,\,g)\in \Z^{\mathrm{pf}}_N\times G_b$
by
\begin{align}
(\tau, \pi): 
\tilde{g} \mapsto (\tau(\tilde{g}),\,\pi(\tilde{g}))\equiv (z,g).
\end{align}
Conversely, an element $(z,g)\in \Z^{\mathrm{pf}}_N\times G_b$
is mapped to an element $\tilde{g}\in\mathcal{G}$ through the map
\begin{align}
\iota \circ s : (z,g) \mapsto \iota(z)\circ s(g)\equiv \tilde{g}.
\end{align}
One verifies that both maps are bijections. In particular, 
$(\tau, \pi)$ is the left inverse of $\iota \circ s$.

We would like to write the group composition rule $\circ$
in $\mathcal{G}$
in terms of the group composition rules $(+,\cdot)$ in 
$\Z^{\mathrm{pf}}_N\times G_b$. To this end, 
we observe for two elements $g,\,h\in G_b$
\begin{subequations}
\begin{align}
\pi(s(g)\circ s(h)) 
=
\pi(s(g))\pi(s(h))
=
g\,h
=
\pi (s(g\,h)),
\end{align} 
where in the last line we have used the fact that
the product $g\,h$ can be written as $\pi(s(g\,h))$.
In other words, the composition rule
\begin{align}
\pi(s(g)\circ s(h))
=
\pi (s(g\,h)),
\end{align}
holds under the projection $\pi$.
However, since $\ker \pi = \mathrm{im}\, \iota$, 
this equation implies
\begin{align}
s(g)\circ s(h)
=
c(g,h)\,
s(g\,h)
\end{align}
\end{subequations}
where $c(g,h)\in \mathrm{im}\, \iota \cong \Z^{\mathrm{pf}}_N$.
In particular, we have the composition rule between the two pairs
$(z,g)$ and $(u,h)$ in $\Z^{\mathrm{pf}}_N\times G_b$
\begin{align}
\label{eq:composition using extension maps}
\iota(z)s(g)\,\iota(u)s(h)
=
\iota(z)\,\iota(\rho_g\,u)\,s(g)\,s(h)
=
\iota(z + \rho_g\,u+c(g,h))\,s(g,h),
\end{align}
where we denote by $\rho_g$ the action of $g$ on the element
$u$ which is due to the possible non-commutativity between
the elements $s(g)$ and $\iota(u)$ in $\mathcal{G}$.

We make the identification $(z,\,g)= \iota(z)s(g) \equiv \tilde{g}$
and $(u,\,h)= \iota(u)s(h) \equiv \tilde{h}$ 
and rewrite Eq.\ \eqref{eq:composition using extension maps}
as the composition rule
\begin{align}
\label{eq:group composition law app}
(z,g)\circ(u,h)=\big(z+\rho_{g}\,u+c(g,h),\,g\,h\big).
\end{align}
Here, $c(g,h)$ is the representative of an extension class $[c]$. 
More precisely, we $c\in C^2(G_b,\Z^{\mathrm{pf}}_N)$
is a 2-cochain such that 
on the one hand the associativity in gorup $\mathcal{G}$
implies that
\begin{subequations}
\begin{align}
\delta^{(2)}_\rho\,c= 0,
\end{align}    
and on the other hand, a group isomorphism $\alpha:\mathcal{G} \to 
\mathcal{G}$ shifts $c$ to $c'$ given by
\begin{align}
c' = c + \delta^{(1)}_\rho\,\alpha.
\end{align}
\end{subequations}
Hence, the inequivalent groups $\mathcal{G}$ that 
can be obtained by extending a group $G_b$
by $\Z^{\mathrm{pf}}_N$ are specified by the 
extension classes $[c]$ in the second cohomology 
group 
$H^{2}\left(G^{\,}_{b},[\Z^{\mathrm{pf}}_{N}]^{\,}_{\rho}\right)$.

\subsection{Group actions on $\Z^{\mathrm{pf}}_N$ are involutive}
\label{appsubsec:group actions are involutive}

The possible group actions $\rho$ are severely restricted.
To derive this restriction, consider an element $g\in G^{\,}_{b}$
and a corresponding element
$\tilde{g}=(z,g)\in\mathcal{G}$.
Consider again the two local operators $A^{\,}_{j}$ and
$B^{\,}_{k}$ with definite parafermion charges
$q^{\,}_{A}$ and $q^{\,}_{B}$, respectively, whose supports
are spatially separated.
They obey the exchange algebra
\eqref{eq:algebra well separated Pf ops}.
Under the action of the symmetry $\tilde{g}$, these operators are
mapped to $A'_{j}$ and $B'_{k}$, respectively.
The group action $\rho_g$ then maps their parafermion charges to
$\rho_g\,q^{\,}_{A}$ and $\rho_g\,q^{\,}_{B}$, respectively.
Consequently, Eq.\ \eqref{eq:algebra well separated Pf ops}
implies that the transformed operators obey
\begin{align}
A'_{j}\,
B'_{k}
=
\w^{\rho^{2}_{g}\,q^{\,}_{A}q^{\,}_{B}
\mathrm{sgn}(k-j)}\,
B'_{k}\,
A'_{j}.
\end{align}
At the time, a unitary transformation must preserve the
exchange algebra of $A^{\,}_{j}$ and $B^{\,}_{k}$,
i.e.,
\begin{align}
A'_{j}\,
B'_{k}
=
\w^{q^{\,}_{A}q^{\,}_{B}\mathrm{sgn}(k-j)}\,
B'_{k}\,
A'_{j}.
\end{align}
Comparing these two exchange relations for operators with
$q^{\,}_{A}=q^{\,}_{B}=1$ yields
\begin{equation}
\label{eq:involutions equation}
\rho^{2}_{g}
=
1\!\mod N.
\end{equation}
Thus, every automorphism in the image of $\rho$ must be
involutive. For $N=p^{k}$, the group of involutive automorphisms of
$\Z^{\mathrm{pf}}_{p^{k}}$ is
\begin{align}
\label{eq:involutive group app}
\mathcal{I}\left(\Z^{\mathrm{pf}}_{p^{k}}\right)
=
\begin{cases}
\Z^{\,}_1, &\hbox{if $p=2$ and $k=1$,}\\
\Z^{\,}_{2}, &\hbox{if $p=2$ and $k=2$,}\\
\Z^{\,}_{2}\times\Z^{\,}_{2},
&\hbox{if $p=2$ and $k>2$,}\\
\Z^{\,}_{2}, &\hbox{if $p>2$.}
\end{cases}
\end{align}
The result for generic $N$ follows from
Eq.\ \eqref{eq:involutive group app} by applying the Chinese
remainder theorem.

\section{Proof of Eq.\ \eqref{eq:general proj rep mu odd}}
\label{appsec:odd parafermion rep identities}

We derive for phases with odd number of parafermions on the boundaries ($\mu=1$) the parametrization \eqref{eq:general proj rep mu odd} of the boundary representation.
Consider without loss of generality an element $\tilde{g} \equiv (0,g) \in \mathcal{G}$ such that there is no charge conjugation, i.e.,  $\rho_g=1$.
Equation\ \eqref{eq:general proj rep mu odd} then reduces to
\begin{equation}
\label{eq:general proj rep mu odd special case}
U_\mathsf{B}(0,g) = V_\mathsf{B}(g) \, \gamma_\infty^{-\kappa(g)},
\end{equation}
where $V_B(g) \in \mathfrak{P}_N^{n_\mathsf{B}}$.
In other words, the auxiliary parafermion may be completely factorized out of $U_\mathsf{B}(0,g)$.
The full parametrization \eqref{eq:general proj rep mu odd} follows from this particular case analogously to the even boundary ($\mu=0$) parametrization \eqref{eq:general proj rep mu even}.

To prove Eq.\ \eqref{eq:general proj rep mu odd special case},
we make the following physical assumption.
Let $Y$ be the generator of the center of the boundary algebra $\mathfrak{P}_N^{n_\mathsf{B}}$ as defined in Eq.\ \eqref{eq:center Y mu odd}.
We ask that the image of $Y$ under the action by conjugation of  $U_\mathsf{B}(0,g)$ must have well-defined (boundary) parafermion charge.

Since $U_\mathsf{B}(0,g)$ is unitary, and thus preserves the algebra's commutation relation, it must map $Y$ to an element of the center.
As $Y$ itself generates the center, we have
\begin{equation}
U_\mathsf{B}(0,g) \, Y \, U_\mathsf{B}(0,g)^\dag = \sum_{a=0}^{N-1} c_a \, Y^a, \quad c_a \in \C.
\end{equation}
Because $Y$ has unit parafermion charge, our assumption implies
\begin{equation}
U_\mathsf{B}(0,g) \, Y \, U_\mathsf{B}(0,g)^\dag = c_a \, Y^a
\end{equation}
for some $a=0,\cdots,N-1$.
Moreover, since $\tilde{g} =(0,g)$ was chosen to realize no charge conjugation, and $Y^N =\mathbb{1}$, we deduce
\begin{equation}
U_\mathsf{B}(0,g) \, Y \, U_\mathsf{B}(0,g)^\dag = \w^{\kappa} \, Y,
\end{equation}
where $\kappa =0,\cdots,N-1$.
Inverting this equation yields
\begin{equation}
Y \, U_\mathsf{B}(0,g) \, Y^\dag = \w^{-\kappa} \, U_\mathsf{B}(0,g).
\end{equation}
Since $Y$ commutes with all parafermion operators on the boundary with the exception of the auxiliary $\gamma_\infty$, this identity can only hold if equation \eqref{eq:general proj rep mu odd special case} is true, where we identified $\kappa \equiv \kappa(g)$.

\twocolumngrid
\bibliography{References}

\end{document}